\documentclass[9pt, sigconf]{acmart}

\usepackage{algorithm}
\usepackage{balance}
\usepackage[noend]{algpseudocode}
\usepackage{caption}
\usepackage{subcaption}
\usepackage{xcolor}
\usepackage{varwidth}
\usepackage{mdwlist}
\usepackage{tabularx}
\usepackage{tikz}
\usetikzlibrary{positioning}
\usepackage{xspace}
\usepackage{url}
\usepackage{titlesec}
\usepackage{subcaption}

\titlespacing{\paragraph}{5pt}{2pt}{2pt}
\titlespacing{\subsection}{1pt}{5pt}{2pt}
\titlespacing{\subsubsection}{1pt}{5pt}{2pt}

\newcommand{\sysname}{ReStore\xspace}

\setcopyright{none}

\begin{document}

\fancyhead{}

\title{\sysname{} - Neural Data Completion for Relational Databases}
\subtitle{[Extended Technical Report]}

\author{Benjamin Hilprecht}
\affiliation{Technical University of Darmstadt}

\author{Carsten Binnig}
\affiliation{Technical University of Darmstadt}

\begin{abstract}
Classical approaches for OLAP assume that the data of all tables is complete.
However, in case of incomplete tables with missing tuples, classical approaches fail since the result of a SQL aggregate query might significantly differ from the results computed on the full dataset.
Today, the only way to deal with missing data is to manually complete the dataset which causes not only high efforts but also requires good statistical skills to determine when a dataset is actually complete.
In this paper, we propose an automated approach for relational data completion called \sysname{}\footnote{This is a technical report of the paper published in SIGMOD 2021.} using a new class of (neural) schema-structured completion models that are able to synthesize data which resembles the missing tuples. As we show in our evaluation, this efficiently helps to reduce the relative error of aggregate queries by up to $390\%$ on real-world data compared to using the incomplete data directly for query answering.
\end{abstract}

\settopmatter{printacmref=false}
\renewcommand\footnotetextcopyrightpermission[1]{}

\newcommand{\hl}[1]{\textcolor{blue}{#1}}
\newenvironment{highlight}{ \color{blue}}{}

\maketitle

\vspace{-1.5ex}
\section{Introduction}
\label{sec:intro}

\paragraph*{Motivation.} 
OLAP and data warehousing play a significant role today for many organizations and enterprises for decision making.
This is evident since many new scalable OLAP services are becoming available in the cloud such as AWS Redshift \cite{redshift}, Snowflake \cite{snowflake}, or Azure data warehousing \cite{azure} that allow customers to analyze large datasets using aggregate queries. 
A critical assumption for OLAP, however, is that the data itself has to be complete before it can be used for decision making, i.e., data in tables is complete and no tuples are missing.
Traditionally, this was achieved by loading data only from well curated (internal) data sources into a data warehouse.
In enterprises, these are typically OLTP systems that store data about customers, products, orders, etc.
However, while data in this context might still require data integration and cleaning \cite{10.1145/3034786.3056124, dong2018data, 10.1145/2588555.2610505, 10.1145/2882903.2912574} since it comes from multiple sources, the data is typically considered complete and all relevant tuples were expected to be present in the data warehouse.

However, this assumption does not hold anymore for many of the more modern analytics scenarios.
Instead of using only well curated (internal) data sources in a data warehouse, more and more external data sources are being used in OLAP scenarios. 
A problem of these external data sources is that the data might be incomplete.
For example, to extend our warehouse we might want to use a CSV file from an open data platform containing information about cities where customers come from --- however, data for some cities is missing.
Moreover, in addition to external data sources there are many more applications where tables can be incomplete such as scenarios where data needs to be collected manually and thus collecting a complete dataset is too expensive or even impossible. 

In case of incomplete tables, classical databases fail since the result of a SQL aggregate query might significantly differ from the results computed on the full dataset which in turn leads to erroneous conclusions in data analysis and decision making.
Moreover, existing techniques that can produce approximate aggregate query answers \cite{chaudhuri2017approximate,agarwal2013blinkdb} on samples might also fail since data is often missing systematically (e.g., samples for some groups in the data are missing completely).
Even worse, the missing data might introduce a bias and hence the data can not be seen as a uniform (random) sample.

For example, suppose we want to create a housing database of rental apartments and their neighborhoods (covering all cities in the US).
While we have a complete neighborhoods table, the apartments data is incomplete since not all states provide this information (i.e., apartments of individual states might be missing completely).
However, this missing data might introduce a bias in the available data, e.g., since most data comes from states with high population densities where rents are higher.
If we now use a SQL aggregate query on the incomplete apartment table to determine the average rental price of apartments across all states, we could obtain (highly) inaccurate results due to the missing apartment tuples. 

\begin{figure*}
	\centering
	\subcaptionbox{Annotated Example Schema.\label{fig:overview:annotated_schema}}[0.33\linewidth]{\includegraphics[width=0.95\linewidth]{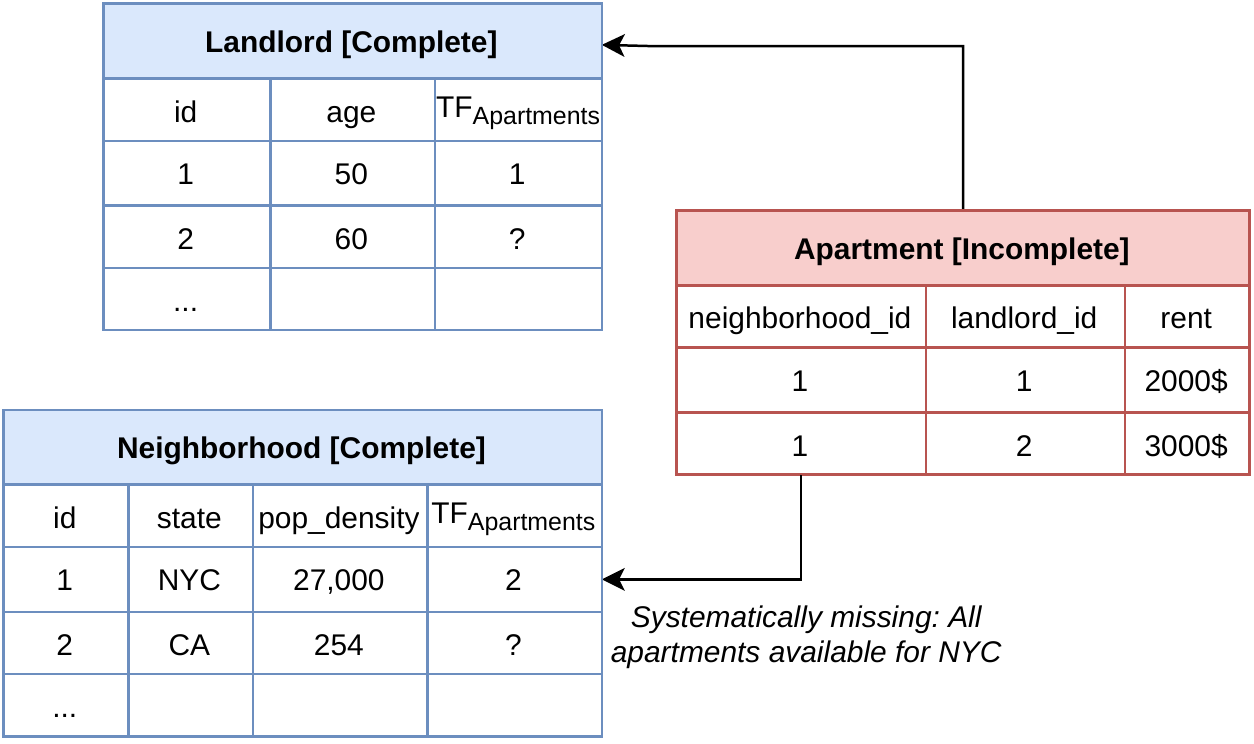}}
	\subcaptionbox{Models Synthesize Missing Tuples.\label{fig:overview:learned_models}}[0.33\linewidth]{\includegraphics[width=0.95\linewidth]{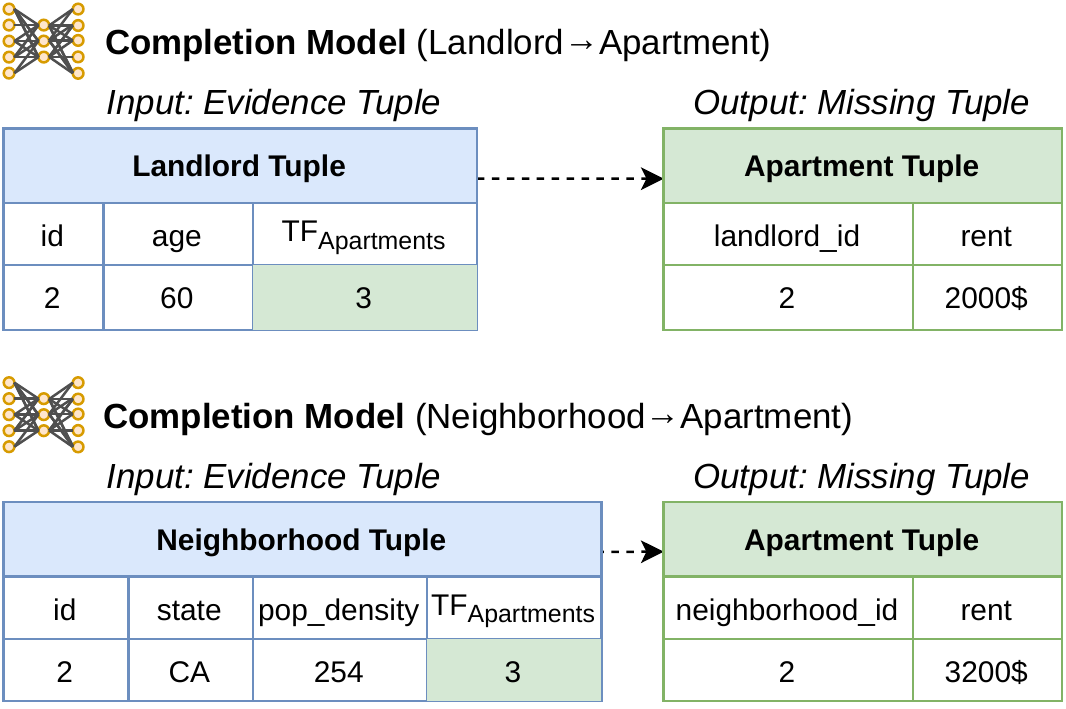}}
	\subcaptionbox{Incompleteness Join.\label{fig:overview:query}}[0.26\linewidth]{\includegraphics[width=0.95\linewidth]{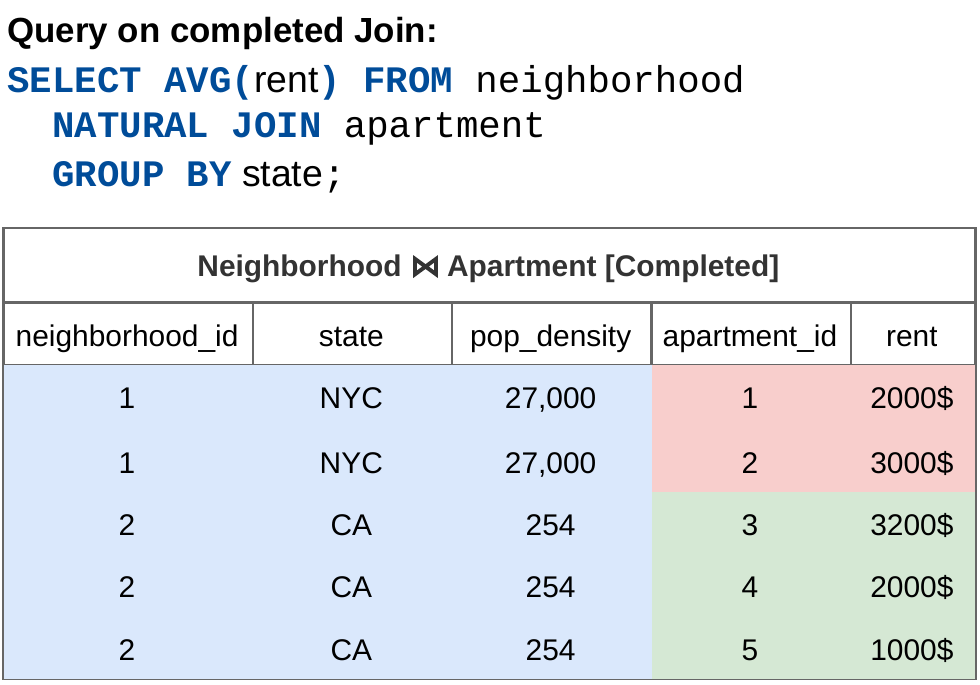}}
	\vspace{-3ex}
	\caption{Overview of \sysname{} to synthesize missing data (green) from existing data (blue and red). (a) Based on the annotated schema and the available data, the completion models are learned. (b) The learned schema-structured model can be used to synthesize a missing \texttt{apartment} tuple using a complete \texttt{neighborhood} tuple as input. (c) The model generates missing data for a given user query at runtime to answer queries over incomplete tables. The generated tuple factors (TFs) allow us to estimate the number of missing tuples.}
	\label{fig:overview}
	\vspace*{-4.5ex}
\end{figure*}

\paragraph*{Contributions.} The only way to deal with missing tuples in databases for OLAP today is to manually add the missing tuples before using the database for decision making.
The manual completion of an incomplete database, however, causes an enormous effort in data acquisition and in checking the completeness of the acquired data.
Moreover, in many situations it might not even be possible to complete a database manually at all. 
In this paper, we thus propose a new learned approach called \sysname{} for automatic data completion for incomplete relational databases.
While there has been already significant work to impute missing values (e.g., replace a missing attribute) including learned approaches \cite{DBLP:journals/pvldb/RekatsinasCIR17, mlsys2020_123, pmlr-v80-yoon18a}, to the best of our knowledge there is no work to synthesize data for incomplete tables in a relational schema where tuples are missing and might introduce a bias.

The main idea of our approach is that we use the complete tables in a database as evidence to synthesize the missing data even if the missing data introduces a bias in the incomplete table. 
For instance, in the example above, we could use the complete neighborhoods table to synthesize the apartment tuples for the missing states.
One might now wonder how the bias from the missing data can be removed.
The intuition is that our neural completion models learn from the available data how typical apartments look like based on information from the neighborhood table (e.g., rents will be higher in neighborhoods with higher population density).
During completion, we take this information from neighborhoods into account to synthesize the missing tuples. 

To enable data completion our approach works in two steps (cf. Figure \ref{fig:overview}): (1) in a first step, the user has to annotate the schema and provide minimal information about the relational dataset once for all queries (such as if a table is complete or incomplete). (2) Once annotated, we learn so called completion models over the incomplete dataset to capture the complex correlations and dependencies across complete and incomplete tables. Using these models, we are then able to synthesize data to complete the missing data for executing aggregate queries. 
As we show in our evaluation, this efficiently helps to reduce the relative error of aggregate queries by up to $390\%$ on real-world data compared to using the incomplete data directly for decision making.

\paragraph*{Outline.} The remainder of the paper is structured as follows.
In Section \ref{sec:overview}, we provide a more formal definition of the problem and present an overview of \sysname{} to tackle this problem.
Afterwards, in Section \ref{sec:schema_models} we present the details of the neural completion models before we then discuss in Section \ref{sec:query_processing} how these models can be used to generate the missing data for answering aggregate queries. 
Furthermore, Section \ref{sec:model_selection} provides further important details on automatic selection of completion models given a user query before we discuss a technique to estimate the confidence of a completion in Section \ref{sec:model_confidence}.
In Section \ref{sec:experiments}, we discuss the results of our evaluation using synthetic and real-world datasets. Finally, we present related work in Section \ref{sec:related_work} and then conclude in Section \ref{sec:conclusion}.

\vspace{-1.5ex}
\section{Overview}
\label{sec:overview}

In this section, we introduce the problem statement before we give an overview of our approach and discuss potential applications and the general assumptions.

\subsection{Problem Statement}
\label{sec:problem-statement}

In brief, the problem that we solve in this paper can be described as follows.
We are given an incomplete database $D^i$ that consists of complete tables $T_1,T_2,\dots$ and incomplete tables $T_j,T_{j+1},\dots$.
The goal is to generate data for the incomplete tables $T_j,T_{j+1},\dots$ based on the available data that allows us to answer a query workload $Q_1,Q_2,\dots,Q_n$ of aggregate queries such that query results on the completed database $Q_i(D^c)$ are close to the query results on the true (complete) database $Q_i(D)$. Note that this formulation allows us to generate missing data individually for each query to answer the given query as accurately as possible. 
However, we can still cache generated data such that we do not need to generate new data for every query individually as we discuss later.

An important question for this problem is how to measure success. Based on our problem definition, a natural metric is how much the relative error of a query result on the incomplete database can be reduced by completing the data; i.e., how much more accurate the query results are after the completion.
The relative error reduction for a given query $Q_i$ can thus be defined as follows: 
\vspace{-0.5ex}
\begin{equation}
\small
\label{eq:error-reduction}
\mathit{Rel.\,Error\,Reduction}=
E_r(Q_i(D^i), Q_i(D)) - E_r(Q_i(D^c), Q_i(D)) 
\vspace{-0.5ex}
\end{equation}
where the relative error $E_r$ is the difference of the two query results normalized by the true query result. 
While for aggregate queries without a group-by, the relative error is trivial, for group-by queries we use the average relative error over all result tuples \cite{hilprecht2020deepdb}.

A limitation of the relative error reduction metric is that it does not show how well the bias of the incomplete database can be reduced independent of a given workload. 
We thus use a second metric called \textit{bias reduction} to measure the success of data completion.
This metric shows how well the true data distribution of a given attribute could be restored. 
For continuous attributes $X,$ the bias reduction is defined as follows:
\vspace{-0.5ex}
\begin{equation}
\small
\label{eq:bias-reduction}
\mathit{Bias\,Reduction}=1-\frac{|\mathit{AVG^c}(X)-\mathit{AVG}(X)|}{|\mathit{AVG}(X)-\mathit{AVG^i}(X)|}
\vspace{-0.5ex}
\end{equation}
where $\mathit{AVG^c}(X), \mathit{AVG^i}(X), \mathit{AVG}(X)$ are the averages of attribute $X$ on $D^c$, $D^i$ and $D$, respectively.
Hence, the bias reduction is normalized in the interval $[0,1]$ where larger values are preferable.
For categorical attributes, we use the fraction of the biased attribute value since an average cannot be computed.

\subsection{Our Approach}
\label{sec:our-approach}

As mentioned before, our approach called \sysname{} to tackle this problem consists of the two steps depicted in Figure \ref{fig:overview}: First, a user has to (once) annotate a database schema before we train neural completion models that can be used to generate the missing data required to execute aggregate queries over the completed database.

\vspace{-0.5ex}
\paragraph{Schema Annotation.}
\label{sec:overview:annotation}
In the annotation step, a user must indicate for a given incomplete database which tables are complete and which ones are incomplete.
An example for an annotated schema is depicted in Figure~\ref{fig:overview:annotated_schema} which consists of three tables of a housing database where two tables are marked as complete (\texttt{landlord} and \texttt{neighborhood}) and one table (\texttt{apartment}) is marked as incomplete.

In addition, information about the relationships between tables needs to be annotated. Here, the user has to provide information whether there are any \textit{complete} foreign-key relationships between tuples from a complete table and an incomplete table.
For example, in Figure~\ref{fig:overview:annotated_schema} all apartments of neighborhoods in NYC are available but not those for CA.
In many of the application scenarios, the information which relationships are complete is known a priori and thus does not cause additional manual annotation overhead.
For example, often a complete subset of data (e.g., apartments of a certain state) is available.

Based on the annotation, so called tuple factors (TF) \cite{hilprecht2020deepdb} can now be automatically computed step to capture information about the relationships across complete and incomplete tables as shown in in Figure~\ref{fig:overview:annotated_schema} (e.g., how many apartments a complete neighborhood has). 
Based on the available data and the computed tuple factors, we then learn our completion models as discussed next.

The user might also have other additional information, which can help to further enhance the quality of the synthesized data. Among these are table sizes for incomplete tables or aggregate statistics (e.g., average rental prices in certain states). Using techniques like iterative proportional fitting \cite{orr2020themis}, this information can be used to improve our generated data. These techniques are orthogonal to our approach and we thus exclude them in the remainder.

\paragraph{Model Training and Data Completion.} 
Given an annotated schema, we can now learn the completion models.
As depicted in Figures~\ref{fig:overview:learned_models}, two completion models have been learned that can either take data from the complete \texttt{neighborhood} table or the complete \texttt{landlord} table to synthesize missing \texttt{apartment} tuples.
By taking complete tables as evidence our models synthesize missing tuples even if there is a bias in the missing data since we capture correlations across tables (e.g., which types of apartments are expected based on the characteristics of the neighborhoods).

These completion models can now be used at runtime to complete the missing data for a given user query.
For instance, if a user wants to know the average rent per state, we first compute the completed join $\texttt{neighborhood}\bowtie\texttt{apartment}$.
More precisely, we introduce a new operator called \textit{incompleteness join} to join complete and incomplete tables that generates the missing tuples needed to make the join complete.
In our example, the incompleteness join would generate apartments for neighborhoods where the data is missing using the appropriate completion model of Figure~\ref{fig:overview:learned_models}.
Once the missing tuples for the join are generated (i.e., the incompleteness join produced its output), we can compute the aggregated result using a normal aggregation operator.

We decided to complete data on a per-query basis at runtime since completing the full database might be too expensive (and actually not needed) for large datasets.
However, it is important to note that the models are not query-dependent and only have to be learned once for an incomplete schema and can be reused across queries.
Moreover, the generated data can still be materialized or even generated a priori as we will discuss in Section~\ref{sec:query_processing}. 

\paragraph{Supported Schema and Queries.} 
In general, our approach supports any relational schema where tables are connected via foreign-key relationships. 
For the workload, we currently limit ourselves to acyclic Select-Project-Aggregate-Join (SPJA) queries where joins are equi-joins along foreign-key relationships which are typical queries for decision making.
An important aspect is that we can support arbitrary filter predicates or aggregate functions as well as any number of group-by attributes. The reason is that once data is completed for a join, we use normal query operators (e.g., filter or aggregate operators) to compute the query results.
Supporting other types of queries, however, is indeed possible.
For example, other join types (e.g., non-equi joins) could be added by deriving tuple factors that represent these join conditions.

\subsection{Application Scenarios}

\paragraph{Systematically Missing Data.} The fact that tuples in databases are missing is often caused by systematic reasons. In other words, the available tuples in a dataset are often not a uniform sample of the full dataset. 
There are many potential reasons for systematically missing data.
For instance, in our housing database, information about apartments might depend on the neighborhood; e.g., in rich neighborhoods landlords are less interested to make the data publicly available.
This induces a bias since the availability of data correlates with properties of the tuples which might lead to wrong conclusions if queries are issued over the incomplete data; e.g., the average rental prices might be underestimated if the missing data is not taken into account.

\paragraph{Integration of Independent Databases.} Another application scenario where our techniques might help is in the integration of data coming from independent sources.
While the individual sources might be complete for the individual purposes they are curated for, incompleteness can still arise when bringing these data sources together.
For example, think of two housing databases, one for the US (West) and one for the US (East).
While the US (West) database might contain three tables with complete data (\texttt{landlord}, \texttt{neighborhood} and \texttt{apartment}), the US (East) database might only contain two tables with complete data (\texttt{landlord} and \texttt{neighborhood} but no apartments).
Hence, in a merged databases for the complete US (West and East), all \texttt{apartment} tuples for US (East) are missing.
With our approach, we could now use the available data from US (West) as evidence to synthesize the missing apartment data for US (East) to get a rough understanding of the housing market based on the information about neighborhoods and landlords.

\paragraph{Expensive Data Collection.} Even in the absence of systematic reasons for missing data it can be a tremendous effort to collect a complete dataset. This is especially true in many data science scenarios. For instance, gathering the data might require extensive surveys or expensive experimental infrastructure for data collection which is especially true for many engineering disciplines or medical use cases. 
In these cases, our approach also helps to reduce the efforts for data collection since often a small sample is sufficient to synthesize the rest of the data.

\subsection{Discussion}
\label{sec:discussion}

The central assumption of our approach is that both missing and available tuples have consistent correlations; i.e., 
while there can be a bias in the available tuples, it is required that the missing tuples have the same correlations between attributes as the remaining tuples. This is not a requirement specifically for \sysname{} but for any system that uses machine learning to complete a dataset since otherwise the available tuples cannot be used as evidence to predict the missing tuples. 
More technically, the conditional distributions of missing tuples $t_m$ given an evidence tuple $t_e$ should be equivalent for remaining and missing tuple distributions, i.e., $P_m(t_m\mid t_e)\approx P_r(t_m\mid t_e)$. If this assumption holds, the main factor determining how accurately the original query result can be restored is the \textit{predictability} of the query attributes as we will later show in our experimental evaluation. If the attributes are not predictable given the evidence tuples, our models will complete the data with lower confidence  (cf. Section~\ref{sec:model_confidence}).

 \vspace{-1.5ex}\section{Learned Completion Models}
\label{sec:schema_models}

A natural fit for the completion task of \sysname{} are so called deep autoregressive (AR) models \cite{germain15made, nash2019autoregressive, papamakarios2017autoregressiveflows}. In the following, we first discuss the relevant background on AR models and then present a first class of simple completion models based on AR models. Afterwards, we present schema-structured autoregressive (SSAR) models which are more expressive than the simple completion models since they can capture the structural information in complex relational schemas that can be used as evidence for generating missing tuples. 

\subsection{Background on Autoregressive Models}
\label{sec:schema_models:ar_background}

Autoregressive models learn a probability distribution by approximating the density of observed variables $p(x_1,\dots,x_n)$.
These models exploit that any density can be decomposed into a product of conditional densities
$p(x)=\prod_{i=1}^{n} p(x_i \mid x_{<i}).$ 
The factors express the conditional density of the $i$-th variable given its predecessors. 

The popular MADE \cite{germain15made} models realize an autoregressive architecture using deep learning techniques. The network obtains a vector $(x_1,\dots,x_n)$ as input and is trained to output the conditional densities $(p(x_1),p(x_2|x_1), \dots, p(x_n| x_{<n}))$. It is ensured that the i-th output $p(x_i \mid x_{<i})$ only depends on inputs with an index $<i$ using masked layers that prevent the flow of information from subsequent inputs. 

Conditional sampling (and hence generating new data) can now easily be implemented using iterative forward sampling. Assume that we are given a partial vector $(x_1, \dots, x_i)$ and want to sample the remaining entries ($x_{i+1},\dots,x_n$) of the vector, i.e., sample from the conditional distribution $p(x_{\ge i}|x_{<i})$. By making use of the autoregressive model, we can first predict the distribution $p(x_{i+1}|x_{\leq i})$ and sample the next variable $x_{i+1}$. We can now repeat the procedure by feeding the vector $(x_1, \dots, x_i, x_{i+1})$ into the network to predict $p(x_{i+2}|x_{\leq i+1})$ and so forth until we have finally computed a conditional sample for all missing variables of the input vector.

\begin{figure*}
	\centering
	\subcaptionbox{Completion Task.\label{fig:model:setup}}[0.25\linewidth]{\includegraphics[width=0.95\linewidth]{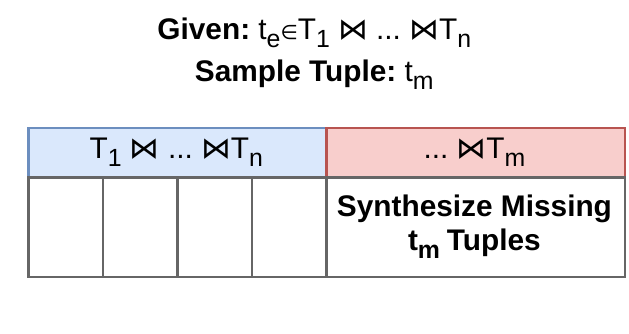}\vspace*{-2.5ex}}
	\subcaptionbox{Simple Completion Model.\label{fig:model:ar_model}}[0.35\linewidth]{\includegraphics[width=0.95\linewidth]{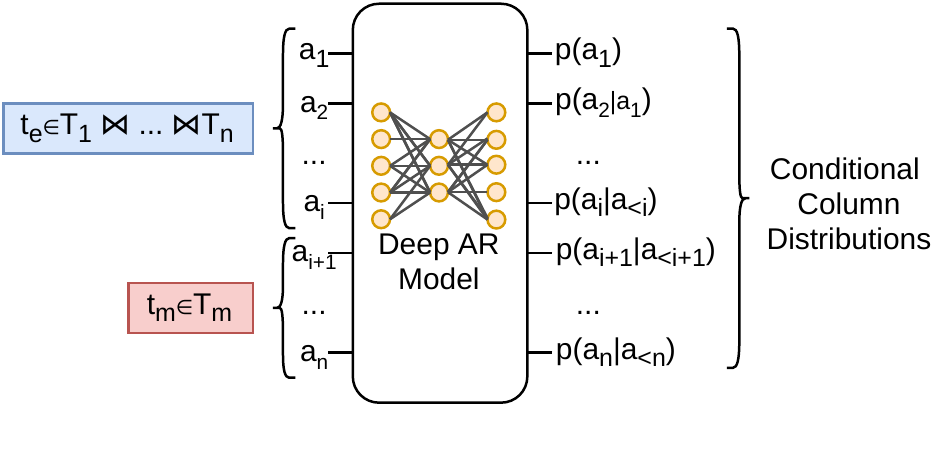}\vspace*{-2.5ex}}
	\subcaptionbox{Schema-Structured Completion Model.\label{fig:model:ssar_model}}[0.34\linewidth]{\includegraphics[width=0.95\linewidth]{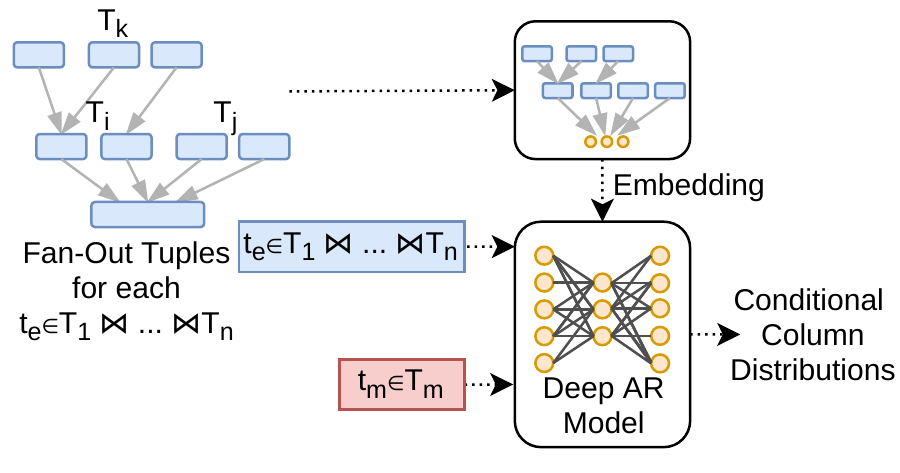}\vspace*{-2.5ex}}
	\vspace{-2.5ex}
	\caption{Learned Completion Models in \sysname{}. (a) The goal is to complete a table $T_{m}$ using the join  $T_{1}\bowtie\dots\bowtie T_{n}$ of complete tables $T_{1}, \dots, T_{n}$ as evidence. (b) Simple completion models are based on autoregressive models and learn conditional distributions $P(a_i|a_{<i})$ over all attributes in $T_{1}\bowtie\dots\bowtie T_{n}\bowtie T_{m}$ (including the incomplete table $T_m$). After learning, we can use conditional sampling to synthesize missing tuples $t_m$ given an evidence tuple $t_e\in T_{1}\bowtie\dots\bowtie T_{n}.$ (c) Schema$-$structured models incorporate additional (so called fan-out) evidence of a tuple $t_e$ using tree embeddings.}
	\label{fig:model}
	\vspace*{-3.5ex}
\end{figure*}

\subsection{Simple Completion Models}
\label{sec:schema_models:ar}

As a first contribution, we present a simple class of completion models based on AR models. 
The general idea of these models is to use tuples of a complete table $t_e\in T_1$ (or of a join of complete tables $t_e\in T_1\bowtie\dots\bowtie T_{n}$) as \textit{evidence} to synthesize a missing tuple of one incomplete table $T_{m}$. 
In other words, the completion models take a tuple $t_e$ as input and synthesize a missing tuple $t_{m}$ for the incomplete table $T_{m}$.

To capture distributions and correlations present in the dataset and eventually generate the missing $T_{m}$ tuples, a completion model for one incomplete table $T_{m}$ is learned over the join of $T_1\bowtie\dots\bowtie T_{n} \bowtie T_m$ (based on the available data).
Clearly, for complex schemata with potentially multiple incomplete tables we need to learn multiple completion models.
An efficient learning procedure for complex schemata is presented at the end of this Section.
In the following, we focus on the question how a single completion model for one incomplete table is derived. 
We first consider the case of using a single complete table as evidence to generate tuples of an incomplete table and later show how joins of tables can be used as evidence.

\paragraph{Single Evidence Table.} 
Let us first consider the case of a single complete table $T_1$ which is connected via a foreign-key relationship to the incomplete table $T_m$. Our goal is to synthesize the missing tuples in $T_m$. To this end, a deep AR model is trained over the join $T_1\bowtie T_m$ (more precisely, all join attributes $a_1,\dots,a_n$ including tuple factors as depicted in Figure~\ref{fig:model:ar_model}) using the available data. Afterwards, for every tuple $t_1$ of the complete table $T_1$ we can synthesize an appropriate tuple for the incomplete table $T_m$ by sampling from the conditional distribution $t_m\sim P(t_m|t_1).$ 
For instance, in our example in Figure \ref{fig:overview} we can synthesize an \texttt{apartment} tuple, given a \texttt{neighborhood} tuple. 

Intuitively, a given \texttt{neighborhood} tuple tells us what a typical apartment in that neighborhood looks like.
Moreover, we can synthesize tuple factors for a given \texttt{neighborhood} tuple if it is not already available. This tells us in addition how many apartments a neighborhood (given its characteristics) has. 
Using the tuple factor, we can now synthesize as many tuples as are missing; e.g., for a neighborhood that should have three apartments but the dataset contains only one, two new \texttt{apartment} tuples need to be synthesized.
This also allows to debias a dataset. 
For instance, the model might predict more missing tuples for neighborhoods in areas with higher population density and since population density and rental prices could be correlated, it will synthesize more expensive apartments resulting in an overall higher average rent.
More details on how more complex queries can be handled an debiased is given in Section~\ref{sec:query_processing}. 
For now, we simply focus on the data generation process for generating one missing tuple $t_{m}$ from a given evidence tuple $t_e$.

\paragraph{Additional Evidence Tables.} Instead of using only a tuple of table $T_1$ as evidence, we can also use information from additional tables $T_2,\dots,T_n$ as evidence.   
The condition for more than one complete table to be used as evidence is that they are connected via foreign-key relationships directly or indirectly to $T_1$. 
This is necessary because otherwise it is not clear which tuples of complete tables should be combined to generate a tuple in the incomplete table. 

For instance, in Figure~\ref{fig:overview}, we cannot use the \texttt{landlord} and the \texttt{neighborhood} tables as evidence in one model to synthesize an \texttt{apartment} tuple. 
While this is technically possible, we do not know a priori in which neighborhoods a landlord has apartments. Trying out all possible combinations is computationally infeasible. Hence, in this particular case, we have to decide whether to use a completion model that uses neighborhoods as evidence or one that uses landlords as input for the completion. This decision is discussed in Section~\ref{sec:model_selection} where we present an algorithm for automatically selecting which data to use as evidence. 
However, as mentioned before, we can still use additional evidence tables $T_2,\dots,T_n$ as long as they are connected via foreign-key relationships to $T_1$ (which  itself is connected to the incomplete table $T_m$). 
The idea is that we can use a tuple from the join $t_e\in T_1\bowtie\dots\bowtie T_n$ as evidence to generate a tuple for $T_{m}$ by sampling from $P(t_{m}|t_e)$. For instance, if the state information of a neighborhood would be represented in a separate table that was connected to the \texttt{neighborhood} table via a foreign-key reference, we could use the joined tuple $t_s \bowtie t_n$ of the \texttt{neighborhood} and \texttt{state} tables as input to more accurately predict a missing \texttt{apartment} tuple $t_a$. 
The attributes of the state table in this case serve as additional features for the deep AR model.

\paragraph{Fan-Out Evidence.} However, even in the case that all additional evidence tables $T_2,\dots,T_n$ are connected to $T_1$ there are limits to which evidence tables can be used. In case one of the tables in $T_2,\dots,T_n$ introduces a fan-out (i.e., the evidence tuple $t_1$ is connected to more than one tuple directly or indirectly in the additional table) the table cannot be used as additional evidence. 
We call this \textit{fan-out evidence}. 
The reason is that if a tuple $t_1$ in $T_1$ has several matching tuples (say in $T_2$), it is not clear which of these tuples should be provided as additional input for the AR model to synthesize a tuple $t_{m}\in T_{m}$. 
For instance, if we had an additional \texttt{school} table in our example which is connected to the \texttt{neighborhood} table, one tuple could have multiple \texttt{school} tuples. To address this issue, we introduce Schema-Structured Completion Models.

\subsection{Schema-Structured Completion Models}
\label{sec:schema_models:ssar}

As mentioned before, simple AR completion models cannot leverage evidence of an additional complete table if it introduces a fan-out. 
This motivates Schema-Structured Autoregressive (SSAR) models which are capable of incorporating this information in the completion process.

\paragraph{Supporting Fan-out Evidence.} Similar to AR models, SSAR models are learned over the join of evidence tables $T_1\bowtie \dots \bowtie T_n$ (which do not introduce any fan-out evidence) and the incomplete table $T_{m}$ as shown in Figure~\ref{fig:model:ssar_model}. 
In order to take the additional tables which introduce a fan-out evidence into account, we perform an acyclic walk on the schema graph. That means for a given evidence tuple $t_e\in T_1\bowtie \dots \bowtie T_n,$ for which we want to generate the missing tuple $t_m,$ we first additionally join tuples from fan-out tables (e.g., $T_i$ and $T_j$ in Figure~\ref{fig:model:ssar_model}). This can be done recursively for tables which have an additional fan-out relationship to tables that are not directly connected to $t_e$ (such as $T_k$ in Figure~\ref{fig:model:ssar_model}).
This results in a tree structure of tuples representing the fan-out evidence, which is then encoded and fed into the neural network in addition to the evidence tuple $t_e$ to predict appropriate tuples $t_m$ of the incomplete table $T_m$.
For instance, for a given \texttt{neighborhood} tuple $t_n$ we would feed the tree with $t_n$ as root and all schools in this neighborhood as children into the model.
To use this tree structure as input to our SSAR models, we encode the tree using a tree embedding architecture. In particular, we use sum-pooling for the child embeddings which are fed into an additional feed-forward network. This architecture was shown to be a universal function approximator for permutation invariant functions \cite{zaheer2017deepsets}. 
We additionally use weight sharing for tuples of the same table to reduce the number of parameters.

\paragraph{Self-Evident Data Completion.} In addition to using tree-structured models to incorporate evidence from additional fan-out tables, we can use tree models also for incorporating the already available data of the incomplete table itself.
For instance, let us again consider the case that the \texttt{apartment} table is incomplete and we wish to complete the join of $\texttt{neighborhood}\bowtie\texttt{apartment}$ using a complete $\texttt{neighborhood}$ table.
Given a \texttt{neighborhood} tuple, the SSAR model has to predict an appropriate missing \texttt{apartment} tuple. As mentioned before, some apartments of a given neighborhood might already be available (but not all). Using tree embeddings, these \texttt{apartment} tuples could also be fed into the SSAR model as additional (self-)evidence. The intuition is that, if there are typical constellations of apartments in a neighborhood (e.g., typically they have comparable prices), this will be learned by the SSAR model and taken into account during the completion further refining the synthesized data. 

\subsection{Learning on Complex Schemata}
\label{sec:schema_models:schema}

So far, we have focused on the question how one individual completion model works.
However, given an annotated schema of a complex database, we have to learn multiple models to potentially synthesize the data for arbitrary joins containing incomplete tables.
More precisely, unless otherwise specified by the user, we want to be able generate tuples for any table $T_x$ using any connected table $T_y$ as evidence. 
Naively, we would have to learn a single SSAR (or AR) model for every every pair of tables $T_x,T_y$ that are connected via a foreign-key to complete $T_x$ using $T_y$ and potentially all other (non fan-out and fan-out) tables connected to $T_y$ that can be used as additional evidence. 
However, this would lead to a high number of models and consequently high training times.
Instead, as we show next models can be merged (before learning them) to reduce the number of models and overall training time significantly.

\paragraph{Merging Example.} For instance, if we want to complete $T_2$ using $T_3$ and $T_1$ using $T_2 \bowtie T_3$ both completions can be done using the same model. We only have to make sure that attributes from $T_3$ are first and that the ones of $T_2$ and $T_1$ are second and third, respectively. This is possible since the model provides both $p(T_1|T_2,T_3)$ and $p(T_2|T_3).$
However, because AR models require a fixed ordering of variables, merging is not always possible. For instance, a model that has to learn $p(T_2|T_1)$ cannot be merged because we cannot find an ordering of variables that allows to predict both $p(T_2|T_1)$ and $p(T_1|T_2,T_3).$

\paragraph{Model Merging.} In our approach, we first require for two models $M_1$ and $M_2$ to be merged that the set of tables of $M_1$ is a subset of the tables of $M_2$ or vice versa. In addition, we have to check whether there exists a consistent variable ordering. 
To this end, we construct a directed graph that contains a node for every involved table. 
For every table that should be completed, we add an arc from every evidence table to this table. Only if the resulting graph is cycle-free a valid ordering of tables can be derived and we merge the models. 
In particular, we use the topological sorting as ordering. We merge models until no more non-conflicting merges are available.
 \vspace{-2.5ex}\section{Query-Driven Data Completion}
\label{sec:query_processing}

In this Section, we show how the completion models (AR and SSAR) can be used to complete data for a given user query that might contain joins over complete and incomplete tables.

\subsection{Overview of Query Processing}
\label{sec:query_processing:qp}

Data completion using \sysname{} happens on a per-query basis at runtime during query processing.
We decided to do the completion on a per-query basis because an offline completion of the full database especially for larger databases is costly and might actually not be required.
As queries, we support SPJA-queries such as the one shown in Figure \ref{fig:overview:query}) that are typical for OLAP with acyclic equi-joins along foreign-keys and arbitrary filters and aggregations (with and without group-by).  

In order to answer such a query, we first compute the join $J = T_{u1} \bowtie \dots \bowtie T_{un}$ over all tables (complete and incomplete) contained in the user query.
During the join computation, we complete the join using our completion models such that $J$ contains all data as if the join would be executed on a complete database.
Afterwards, we then apply filter predicates, aggregations and groupings to answer the user query over the completed join.

For efficiency, we push down filter predicates and generate only missing data for the requested subset of tuples in the join of the query.
However, for simplicity of explanation, we assume in the following that filters are executed after the join.
Moreover, as we describe Section~\ref{sec:query_processing:optimizations}, this also enables optimizations to reuse the generated data for subsequent queries.

\subsection{Single Incomplete Table in a Query}
\label{sec:query_processing:stables}

We now first consider the case where a single table $T_m$ in the join $T_{u1} \bowtie \dots \bowtie T_{un}$ of the user query is incomplete (as it is the case in Figure \ref{fig:overview:query}) and discuss the case where multiple tables in the user query are incomplete later.
In principle, different models could be available to synthesize data for the incomplete table $T_m$.
We now discuss how a completion works if a model $M$ is already selected and discuss in Section \ref{sec:model_selection} how to select a completion model.

Moreover, we initially assume that the tables that are used as evidence for generating missing data for $T_m$ are among the complete tables in the join $T_{u1} \bowtie \dots \bowtie T_{un-1} \bowtie T_{un}$.
To differentiate in the sequel between the tables  $T_1 \bowtie \dots \bowtie T_n$ needed as evidence for the completion model $M$ and the user join $T_{u1} \bowtie \dots \bowtie T_{un}$, we use the terms \textit{completion path} and \textit{query path}, respectively.

In the following, without loss of generality, we assume that $T_m = T_{un}$ is the incomplete table and $T_m$ is connected to the complete (evidence) table $T_{u1}$ via a foreign key or vice versa.
The step of extending a join of complete tables $T_{u1} \bowtie \dots \bowtie T_{un-1}$ with an incomplete table $T_{un}$ to $T_{u1} \bowtie \dots \bowtie T_{un-1} \bowtie T_{un}$ while generating the missing tuples is called \textit{incompleteness join}.

\paragraph{Completion Path equals Query Path.} 

The simplest case for an incompleteness join is where the query path is equal to the completion path.
Imagine, there is one more \texttt{state} table in our example of Figure \ref{fig:overview} which has a reference to the \texttt{neighborhood} table and a user requests a join of the complete \texttt{state} and \texttt{neighborhood} tables with the incomplete \texttt{apartment} table. In this case, we could use a completion model that uses states and neighborhoods as evidence to synthesize apartments, the query path and completion path would be both $\texttt{state} \bowtie \texttt{neighborhood} \bowtie \texttt{apartment}$.

For executing an incompleteness join in this case, we first join the complete evidence tables $T_e = T_{u1} \bowtie \dots \bowtie T_{un-1}$ (\texttt{state} and \texttt{neighborhood} in our example). 
Afterwards, we iterate over all evidence tuples $t_e \in T_e$ and synthesize the missing data for the user join. 
In case of SSAR models additional fan-out evidence tables need to be joined separately to construct the query tree for each evidence tuple $t_e$ which is fed into the SSAR model.
For generating the missing data using the completion model, we have to differentiate whether the relationship of $t_e$ and tuples of the incomplete table $T_m$ is a \texttt{1:n} or \texttt{n:1} relationship (i.e., if one evidence tuple $t_e$ has multiple join partners or one join partner in the incomplete table). 

In case of a \texttt{1:n} relationship, we first have to determine how many $t_m$ tuples have to be generated per $t_e$ tuple which can be estimated using the tuple factors which are learned by the corresponding AR or SSAR completion model. 
Moreover, we have to determine how many $t_m$-tuples already exist (since some might already be available but not all) and 
synthesize only the missing number of tuples. 
This can be done efficiently during joining by first creating a hash-map on the incomplete table (which is needed for joining anyway) that additionally counts the occurrences of tuples with the same foreign-key in the $T_m$ table. 
For instance, if we want to synthesize apartments given the join $\texttt{neighborhood}\bowtie\texttt{states}$, we first have to predict how many apartments we expect to see per neighborhood, i.e., the tuple factor per neighborhood. Afterwards, we synthesize the appropriate number of apartments using the join of \texttt{state} and \texttt{neighborhood} as evidence. 
For the output of the incompleteness join, we then join $t_e$ with all the existing and synthesized tuples.

In case of a \texttt{n:1} relationship, we can disregard tuple factors and only need to generate one missing tuple $t_m$ per evidence tuple $t_e$ if needed.
For example, for a join of the \texttt{landlord} table with the incomplete \texttt{apartment} table, we synthesize a landlord only for apartments where the \texttt{landlord} tuple is missing.

\begin{figure}
	\centering
	\includegraphics[width=0.99\linewidth]{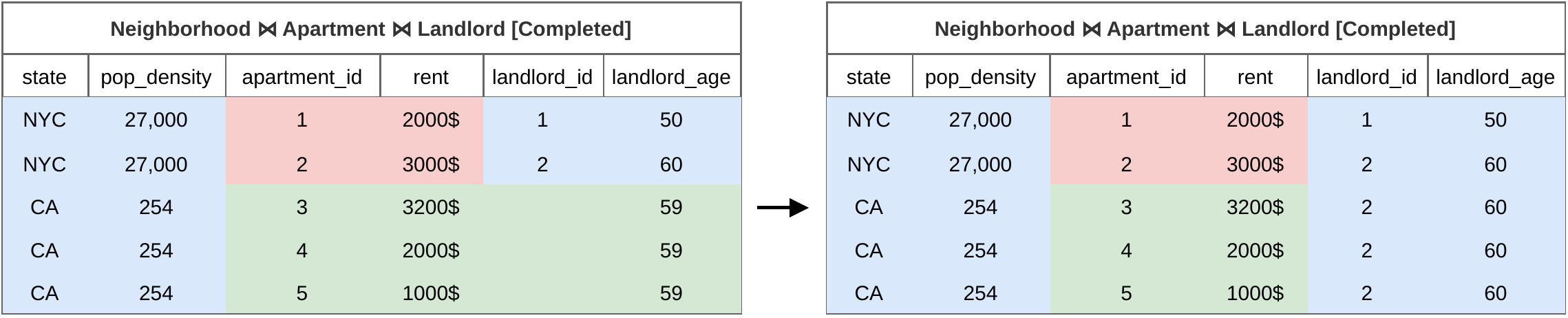}
	\vspace{-2.5ex}
	\caption{Nearest Neighbor Replacement. Foreign-keys are not synthesized for the \texttt{apartment} table and thus the tuples cannot be joined with the complete \texttt{landlord} table. Hence, \texttt{landlord} tuples are first synthesized and afterwards replaced with ``similar'' \texttt{landlord} tuples}
	\label{fig:ann}
	\vspace*{-4.5ex}
\end{figure}

\paragraph{Completion Path contained in Query Path.}

We now consider the case that the completion path is contained in the query path (i.e., the query path contains more tables than the completion path).

In this case, we use a similar approach as before and use the completion path tables as evidence for the model to generate the missing tuples in $T_m$ but then need to join the remaining complete tables of the user query (not in the completion path). 
For instance, assume a user wants to join all three tables in the example in Figure \ref{fig:overview} (\texttt{neighborhood}, \texttt{apartment}, and \texttt{landlord}).
To process such as query, we could use a completion model which allows us to generate \texttt{apartment} tuples from \texttt{neighborhood} tuples to produce a ``completed'' join for those two tables.
Afterwards, we then need to join this output with the complete \texttt{landlord} table.
However, our completion models do not generate foreign keys (to the landlords) for the synthesized \texttt{apartment} tuples since AR and SSAR models are not suited for generating such type of information.

Hence, we cannot use a normal join operator for joining the output of an incompleteness join with the next complete table (e.g., with \texttt{landlord} in our example) but have to process this join in a different manner.
In this case, we again use a completion model that allows us to generate a new \texttt{landlord} tuple using apartments and a neighborhood as evidence as depicted in Figure~\ref{fig:ann} (left).
Since the \texttt{landlord} table, however, is a complete table we then replace the synthesized tuple with an existing tuple that has the highest similarity (i.e., lowest euclidean distance) with the synthesized tuple. For instance in Figure~\ref{fig:ann}, the last three synthesized \texttt{landlord} tuples are replaced with the second landlord from the complete table since they are very similar.

However, an \textit{exact} nearest neighbor replacement of the generated \texttt{landlord} tuple would come at a high cost of computing the pairwise distances of all synthesized tuples and tuples of the complete table during query processing. Hence, we employ approximate nearest neighbor approaches and batching for the replacement. This is crucial to achieve a competitive performance.
In general, this join procedure has to be used if foreign keys in an intermediate result are missing but required for a join with a complete table. Otherwise, normal joins can be used. Although the synthesized data is of high-quality the replacement is required to fully comply with the user annotations - it is unexpected to see new synthesized tuples for complete tables.

\subsection{Multiple Incomplete Tables in a Query}
\label{sec:query_processing:mtables}

We have now discussed all techniques required to complete a user query where the query path includes only a single incomplete table. The case of several incomplete tables can now easily be derived. In particular, we again assume that the completion path is given and repeatedly apply incompleteness joins as before and use the nearest neighbor replacement where appropriate.
The order which table to complete first is determined using the techniques in Section \ref{sec:model_selection} to automatically select the best completion model.

There is only one difference compared to the single incomplete table case since we have to apply the nearest neighbor replacement also for incomplete tables. In particular, if we synthesize tuples for an incomplete table, we might still synthesize too many tuples since foreign-keys of previous tables might not be generated and thus even though the tuples are still in the database, they would still not appear in the resulting join. Hence, we have to estimate how often a tuple of the incomplete table should appear in the full join and complete accordingly.

The pseudocode for the general case which summarizes the discussions in Sections \ref{sec:query_processing:stables} and \ref{sec:query_processing:mtables} is shown in Algorithm~\ref{alg:single_path}.

\begin{algorithm}[!t]
	\scriptsize
	\caption{Single Table Completion}\label{alg:single_path}
	
	\begin{algorithmic}[1]
		\Require Requested Join Tables $\mathbf{J}_{\mathit{req}}=T_{u1},\dots,T_{un}$
		\Require $T_{1},\dots,T_{n}$ (Path from complete Table $T_1$ to $\mathbf{J}_{\mathit{req}}$)
		\Ensure Approximated Complete Join $T_{u1}\bowtie\dots\bowtie T_{un}$
		
		\State $\mathbf{J}\leftarrow T_1$
		\For{$T_i$ \textbf{in} $T_1,\dots,T_{n-1}$}
		\State \textit{// Incompleteness Join} \label{alg:single_path:join_processing}
		\State $\mathbf{J}_{\mathit{incomplete}} \leftarrow J \bowtie T_i$
		\If{$T_i \bowtie T_{i+1}$ is Fan-Out}
		\State Predict Tuple Factor $\mathcal{F}_{T_{i+1}\leftarrow T_i}$ for every $t\in \mathbf{J}$
		\State $\mathcal{F}_{T_{i+1}\leftarrow T_i}$ $\leftarrow$ $\mathcal{F}_{T_{i+1}\leftarrow T_i}$ - Current No of Join Partners in $T_{i+1}$
		\State $\mathbf{J_{\mathit{syn}}} \leftarrow$ Duplicate each $t\in J$ $\mathcal{F}_{T_{i+1}\leftarrow T_i}$ times
		\Else
		\State $\mathbf{J_{\mathit{syn}}}$ Tuples in $\mathbf{J}$ without Join Partner in $T_{i+1}$
		\EndIf 
		
		\State
		\State \textit{// AR or SSAR Tuple Synthesis}
		\State $\mathbf{M}\leftarrow$ Completion Model for $T_i\rightarrow T_{i+1}$
		\State $\mathbf{J_{\mathit{syn}}} \leftarrow$ Synthesize Columns of $T_{n+1}$ in $\mathbf{J_{\mathit{syn}}}$ using $\mathbf{M}$ \label{alg:single_path:join_processing_end}
		
		\State
		\State \textit{// Euclidean Replacement}
		\If{Last Join or Next Join Fan-Out}
		\State $\mathbf{J}_{\mathit{syn}}\leftarrow$ \texttt{euclidean\_replace}($\mathbf{J}_{\mathit{syn}}$, $T_{i+1}$)
		\EndIf
		
		\State $\mathbf{J} \leftarrow \mathbf{J}_{\mathit{syn}} \cup \mathbf{J}_{\mathit{incomplete}}$
		
		\EndFor
\State \Return $\mathbf{J}$
	\end{algorithmic}
\end{algorithm}

\subsection{Additional Cases for Data Completion}

\paragraph{Completion with Additional Tables.}
We have now considered the case of incomplete tables in a user query under the condition that the completion path is a subset of the requested query path. 
However, this is not necessarily the case since the completion path can also contain additional tables: for instance, if the user queries the \texttt{landlord} and the \texttt{apartment} table but for the completion of the \texttt{apartment} table the lower model in Figure~\ref{fig:overview:learned_models} is chosen which uses neighborhoods as evidence.
The high-level idea for query processing in such a case is that we first use the join over all tables in the completion path to synthesize the missing data for the incomplete table (e.g., the \texttt{apartment} table is completed using the \texttt{neighborhood} table) and afterwards potentially have to reweight tuples according to the introduced fan-out similar to \cite{hilprecht2020deepdb}.

\paragraph{Multi-Path Completion.} Another interesting case is that using only a single path for the completion of one incomplete table can be insufficient.
For instance, let us consider a slightly modified schema of a complete \texttt{apartment} table , an incomplete \texttt{neighborhood} table and an additional complete \texttt{school} table which has a foreign-key relationship to the neighborhoods. 
If a user now simply queries the \texttt{neighborhood} table and we complete the neighborhoods via the \texttt{school} table, neighborhoods that do not have any schools will be missing (since we never generate them if we use a completion path from \texttt{school} to \texttt{neighborhood}). 
In these cases, we use all paths to synthesize data and combine data based on tuple factors.

\subsection{Further Optimizations}
\label{sec:query_processing:optimizations} 

While our data completion process synthesizes data at query runtime, data which is synthesized for one query can be reused for related queries. This allows for (i) caching of data synthesized at runtime or (ii) an offline completion independently of the workload.

We first discuss how data for completed joins can be reused. 
In particular, since aggregations and filters are applied after completing a join to approximate a query $Q$ in \sysname{}, the completed data of $Q$ can be reused for a query $Q'$ if they use the same join path $J$. 
Moreover, if a query $Q'$ requires additional tables not covered in $J$, we can start from $J$ and generate additional data incrementally for further incomplete tables.
Finally, if $Q'$ only requires a sub-path of $J$, we can reuse the data by projecting $J$ to the tables required by $Q$. 

Second, as mentioned before we can also generate missing data prior to the query runtime. One way is to predict which queries will occur at runtime and thus optimize which incompleteness joins to create. However, if there is no knowledge about potential queries, simple heuristics-driven strategies can be used. In particular, we can create data for every pair of a joinable incomplete and complete table. This would allow us to answer any query on a single incomplete table or a join of a complete and incomplete table without the need to generate data.

\begin{figure*}
	\centering
	\subcaptionbox{Housing Schema.\label{fig:datasets:housing}}[0.15\linewidth]{\includegraphics[width=0.51\linewidth]{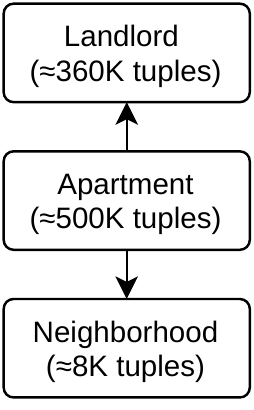}}
	\subcaptionbox{Movie Schema.\label{fig:datasets:movies}}[0.18\linewidth]{\includegraphics[width=0.95\linewidth]{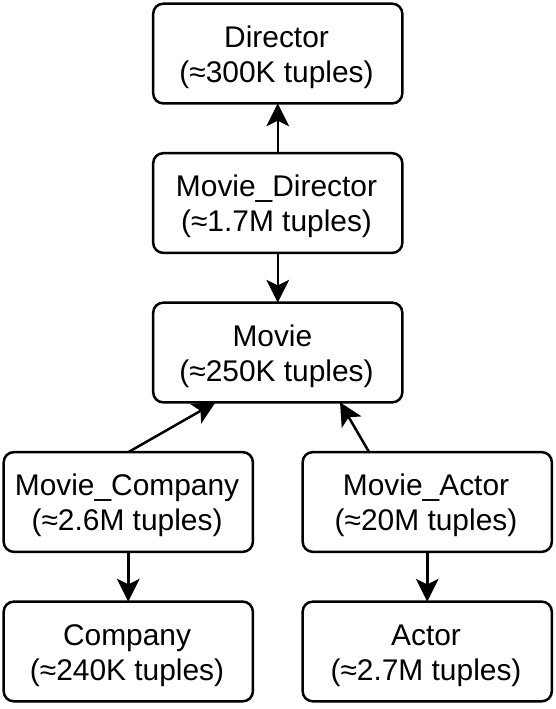}}
	\subcaptionbox{Completion Setups.\label{fig:datasets:setups}}[0.6\linewidth]{\scriptsize
		\centering
		\begin{tabular}{lllllll}\toprule
			& & Tuple Factor & Keep Rates & & & \\
			Setup & Biased Attribute & Keep Rate & \texttt{landlord} & \texttt{apartment} & \texttt{neighborhood} &  \\\midrule
			$H_1$ & \texttt{apartment.price} & 30\% & 100\% & 20-80\% & 100\% & \\
			$H_2$ & \texttt{apartment.room\_type} & 30\% & 100\% & 20-80\% & 100\% & \\
			$H_3$ & \texttt{apartment.property\_type} & 30\% & 100\% & 20-80\% & 100\% & \\
			$H_4$ & \texttt{landlord.landlord\_since} & 30\% & 20-80\% & 100\% & 100\% & \\
			$H_5$ & \texttt{landlord.response\_rate} & 30\% & 20-80\% & 100\% & 100\% & \\\midrule
			Setup & Biased Attribute & & \texttt{movie} & \texttt{director} & \texttt{actor} & \texttt{company} \\\midrule
			$M_1$ & \texttt{movie.production\_year} & 20\% & 20-80\% & 100\% & 100\% & 100\% \\
			$M_2$ & \texttt{movie.genre} & 20\% & 20-80\% & 100\% & 100\% & 100\% \\
			$M_3$ & \texttt{movie.country} & 20\% & 20-80\% & 100\% & 100\% & 100\% \\
			$M_4$ & \texttt{director.birth\_year} & 20\% & 80\% & 20-80\% & 100\% & 100\% \\
			$M_5$ & \texttt{company.country\_code} & 20\% & 80\% & 100\% & 100\% & 20-80\% \\
			\bottomrule
	\end{tabular}}
	\vspace{-3.5ex}
	\caption{Datasets and Completion Setups.}
	\vspace{-2.5ex}
	\label{fig:datasets}
\end{figure*}

\vspace{-2.5ex}\section{Model and Path Selection}
\label{sec:model_selection}

In the approach discussed so far there are some degrees of freedom. In particular, whether we should rather learn AR or SSAR models and which complete tables (i.e., which completion path) should be used for the completion. Both decisions can have a significant impact on the quality of the completion. Intuitively, while the first aspect determines whether we learn a model that is fitting the data well, the second aspect is important because different completion tables have a varying significance for the join we want to complete.

\paragraph{Basic Selection.} To decide whether a model should be used for completion of an incomplete table (or not), it is important to check the accuracy (i.e., test loss) of the models prior to using the model for completion. If the accuracy is too low this means that the true attribute values can hardly be reconstructed since they are not predictable and the bias is likely not reduced significantly.

\paragraph{Advanced Selection.} For the remaining models we have to estimate the quality of each completion model. 
To this end, we derive additional incomplete scenarios with the given incomplete dataset as ground truth to assess model and path quality. 
The underlying assumption is that if the models and paths are able to reconstruct our incomplete dataset they are also able to perform the actual completion with high accuracy.

In practice, the user often suspects a bias in the data but the extent of it is unclear. 
This information can additionally be provided by the user and used for the model selection. 
For instance, an incomplete table might cover more high-population neighborhoods and thus the user expects an overestimation of the average rent.
As we will show in our experiments, this additional information can significantly improve the quality of the synthesized data.

\vspace{-1.5ex}
\section{Completion Confidence}
\label{sec:model_confidence}

It is crucial for practitioners to be aware of the confidence of query results after the data completion. For this, we provide confidence interval estimations of the query results for how certain our models are when synthesizing missing data. 
In the following, we start with the simple case that involves only a single incomplete table and then explain the more general case.

\subsection{Simple Case}

For the simple case, we assume that we have a similar housing database as before but with only two tables: an incomplete apartments table where apartments can have two types (large and small) and a complete neighborhoods table.
Furthermore, assume that a user issues a count-query that joins these two tables to compute the frequency of the two apartment types for which we want to compute confidence intervals. 
Intuitively, if the neighborhood tuples do not provide strong evidence about the types of missing apartments (i.e., if there is a low correlation), the completion models will predict both apartment types with equal probabilities for each missing tuple. In this case, we should have a low confidence and predict wide confidence intervals. In contrast, if the model predicts the apartment type with high certainty, the confidence interval should be more tight. 

In order to compute confidence intervals for a query over an incomplete table, we use the following two-step procedure: (1) We first compute the certainty $C(t_e)$ of a prediction for an attribute of a missing tuple given an evidence tuple $t_e$ in \sysname{}. For this, we compare the probability distribution of the predicted attribute value $P_{\mathit{model}}$ for one synthesized tuple with the distribution of the attribute values in the training data $P_{\mathit{incomplete}}$. If the model is uncertain when synthesizing an attribute value for one missing tuple $t_m$, given the evidence tuple $t_e$, it will simply predict the distribution of values in the training data (i.e., $P_{\mathit{model}}\approx P_{\mathit{incomplete}}$).
However, if the model is certain given an evidence tuple, it will predict a particular attribute value (e.g., a large or a small apartment type) with higher probability. Hence, for computing the certainty of a prediction, we compute the similarity of the distribution $P_{\mathit{model}}$ with $P_{\mathit{incomplete}}$ using the KL-divergence and normalize it to $[0,1]$ by $1-\mathit{exp}(-D_{\mathit{KL}})$.
(2) Second, we compute confidence intervals for each synthesized tuple as follows. For this, we introduce a lower and upper bound distribution ($P_{\mathit{lower}}$ and $P_{\mathit{upper}}$). In our example, we use a distribution for the upper bound $P_{\mathit{upper}}$  where one particular apartment type (e.g., the small apartments) occurs in 95\% of the cases (for a 95\%  confidence). The upper bound of our confidence intervals can then be computed using $C(t_e) P_{\mathit{model}}(t_e) + (1-C(t_e))P_{\mathit{upper}}$. For the lower confidence interval, we simply replace $P_{\mathit{upper}}$ by $P_{\mathit{lower}}$ where $P_{\mathit{lower}}$ represents the distribution where apartments only occur in 5\%.

\subsection{General Case} 

The procedure above can be generalized to queries that (1) involve multiple incomplete tables and (2) other aggregate functions. 
In order to support (1), we generate the missing tuples using the completion models similar to the the normal completion process. However, for every query attribute that has to be synthesized we define an individual distribution $P_{\mathit{lower}}$ and $P_{\mathit{upper}}$ (based on the given confidence) and compute the model confidence intervals as described before. Again, instead of using $P_{\mathit{model}}$ directly, we use $C(t_e) P_{\mathit{model}}(t_e) + (1-C(t_e))P_{\mathit{lower}}$ for the synthesized attributes when computing the lower bound and similarly $P_{\mathit{upper}}$ for the upper bound. 
For this process, we assume that attribute values of different tables are correlated to generate conservative (i.e., worst case) confidence bounds.
(2) As mentioned before we can also support other aggregate functions.
For example, to support average in addition to count aggregates we define $P_{\mathit{lower}}$ and $P_{\mathit{upper}}$ for continuous attributes. Moreover, sum aggregates can be treated as a combination of average and count.
Note that we currently only support completion confidence intervals for query attributes used in an aggregation (i.e., count, avg, sum). For other query types, we can resort to per-query statistics that we show a user such as the ratio of synthesized vs. existing tuples.

 \vspace{-2.5ex}\section{Experimental Evaluation}
\label{sec:experiments}

In this Section, we evaluate both the quality of the completed relational datasets as well as several performance aspects of \sysname{}:\footnote{Code and data is available online: \url{https://github.com/DataManagementLab/restore}} \emph{(Exp.~1 \& 2) Data Completion:} We first evaluate how well our models can correct incomplete datasets given certain data characteristics. \emph{(Exp.~3) Query Processing:} In addition, we demonstrate the end-to-end accuracy of our approach using aggregate queries on real-world datasets. 
\emph{(Exp.~4) Accuracy and Performance:} We finally discuss the accuracies of the different models and the model selection as well as the time required for model training and data completion.

\begin{figure*}
	\centering
	\subcaptionbox{Bias Reductions.\label{fig:syn:red}}[0.67\linewidth]{\includegraphics[width=0.99\linewidth]{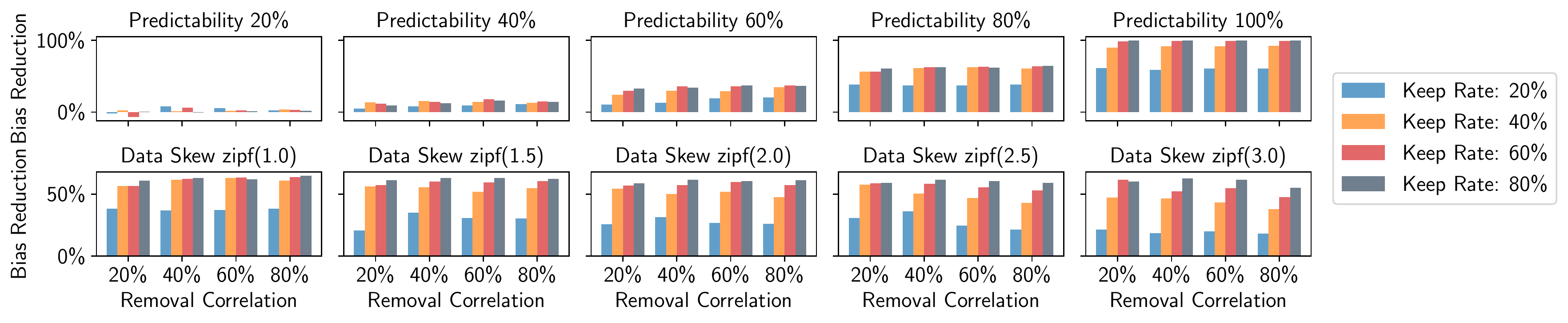}\vspace*{-2ex}}
	\subcaptionbox{Training Loss.\label{fig:syn:loss}}[0.14\linewidth]{\includegraphics[width=0.99\linewidth]{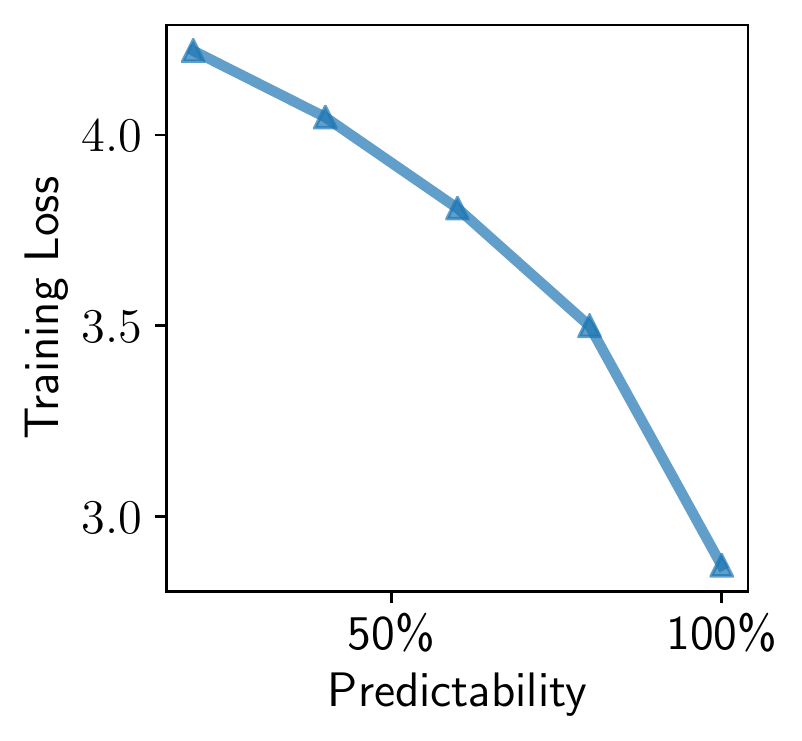}\vspace*{-2ex}}
	\subcaptionbox{Fan-out Predictability.\label{fig:syn:ssar_fan_out}}[0.18\linewidth]{\includegraphics[width=0.86\linewidth]{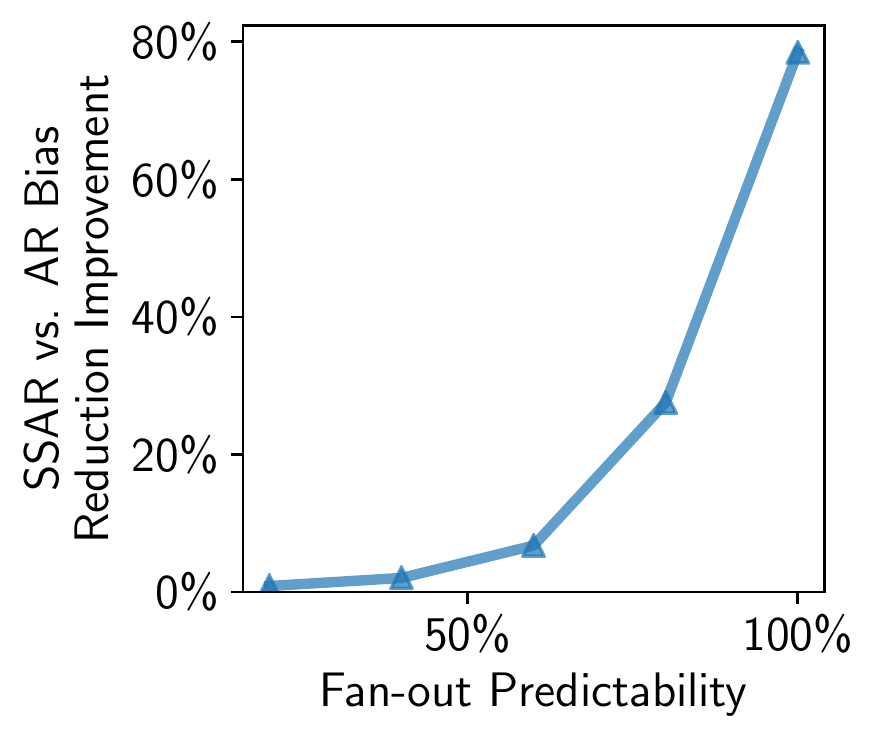}\vspace*{-2ex}}
	\vspace{-3.5ex}
	\caption{Bias Reductions for the Synthetic Datasets. (a) Predictability affects reconstructability. Skewness has no effect on data completion. (b) The test loss is an effective criterion for model selection as discussed in Section \ref{sec:model_selection}. (c) SSAR models are superior over AR models since they can capture fan-out evidence (called fan-out predictability in the Figure).}
	\vspace{-2.5ex}
	\label{fig:syn}
\end{figure*}

\subsection{Datasets and Implementation}

\paragraph{Datasets.} We first evaluate our approach on a synthetic dataset to investigate which factors determine the quality of our completion in isolation.
However, restricting ourselves to synthetic datasets is insufficient since they do not exhibit as complex distributions and correlations as real-world datasets. We thus also evaluate our approach on two real-world relational datasets with different complexity. The first schema is a housing dataset derived from the Airbnb data\footnote{\url{https://public.opendatasoft.com/explore/dataset/airbnb-listings}} which we normalized to obtain different relations for landlords, neighborhoods and apartments (Figure~\ref{fig:datasets:housing}). The movies schema is derived from the popular IMDB\footnote{\url{http://homepages.cwi.nl/~boncz/job/imdb.tgz}} dataset but with two important differentiations. We first merged the \texttt{movie\_info} table information into the \texttt{movie} table to obtain more interesting attributes, i.e., genre and rating. Moreover, we explicitly divided the \texttt{person} relation into actors and directors exhibiting a more interesting relational structure as depicted in Figure~\ref{fig:datasets:movies}. 
For both datasets we create incomplete versions by removing a varying ratio of tuples to simulate different degrees of incompleteness.
Details on how we removed data will be given in our experiments.

\paragraph{Implementation.}
All models were implemented with PyTorch \cite{NEURIPS2019_9015}. For the AR models, we used the model in \cite{yang2019naru} as a starting point.\footnote{\url{https://github.com/naru-project/naru}} Similar to \cite{yang2019naru}, we use learned embeddings to represent attribute values in the AR and SSAR completion models. In particular, we use the MADE \cite{germain15made} architecture for AR models with residual connections and ReLU activation functions. For the neural tree architectures in the SSAR models we use a deep sets architecture \cite{zaheer2017deepsets}. 

\subsection{Exp. 1: Data Completion on Synthetic Data}
\label{sec:experiments:exp0}

In this experiment, we first study the factors that determine the quality of our completions.
For this, we generate different synthetic datasets where we vary different data characteristics that might influence how well the data is reconstructable.
As an additional sanity check, we want to investigate if our automatic model and path selection strategies are able to identify cases that prevent the data completion.
We first introduce the metrics and setup before we discuss our results on synthetic data.

\paragraph{Completion Setups.} 
For this experiment we use a simple synthetic dataset with only two tables: a complete table $T_A$ with a single attribute $A$ and an incomplete table $T_B$ with a single attribute $B$ where $T_B$ has a foreign-key relationship to $T_A$. 
As main parameters which might have an influence on how well a dataset is reconstructable we vary the \textit{predictability} (i.e., how well an attribute can be estimated) and the \textit{skew}.
In particular, the categorical attribute $B$ is generated such that $B$ can be perfectly predicted given $A$ (i.e., $B$ is functional dependent on $A$) and we then incrementally add more noise to reduce the predictability. 
Moreover, the attribute $A$ is generated either using a uniform or skewed distribution where the Zipf factor is varied (for a fixed predictability of 80\%). In addition, we not only vary the predictability of $B$ given $A$ but also the \textit{fan-out predictability} (i.e., how well a missing tuple in $T_A$ can be predicted given other $T_A$ tuples).

In order to derive an incomplete dataset from the synthetic dataset, we systematically remove tuples using two parameters: \textit{removal correlation} and \textit{keep rate}. The keep rate determines the percentage of tuples which are not removed from table $T_B$. In order to introduce a bias, we correlate the probability of a tuple being removed with the value of the attribute $B$. The corresponding parameter controls the strength of this correlation. In particular, we correlate the removal probability with the appearance of one attribute value of $b \in B.$

\paragraph{Metrics and Baselines.} We use the metrics defined in Section \ref{sec:problem-statement}.
For evaluating the quality of data completion (Exp. 1 and Exp 2), we show the \textit{bias reduction} since it is independent of a given workload. 
For experiments (Exp. 3) which involve a workload, we additionally show the \textit{relative error}. Unless otherwise stated, we report the metrics for an optimal model and path selection. We provide a dedicated analysis of the model and path selection in Exp. 4. Both metrics show how well we can reconstruct the complete (true) dataset compared to using the incomplete dataset. We do not compare to other baselines, since to the best of our knowledge no approach exists that is capable of completing relational datasets across tables.

\paragraph{Results.} As we see from Figure~\ref{fig:syn:red} (upper row) the predictability is the key factor determining the success of the debiasing. Intuitively, a high predictability allows our model to accurately estimate the missing values of the attribute $B$ for the missing tuples. However, in cases where the attribute $B$ cannot accurately be predicted given attribute $A$, the test loss of the model is also higher as shown in Figure~\ref{fig:syn:loss}. This confirms that checking the model accuracy is an effective criterion for model selection as discussed in Section~\ref{sec:model_selection}. In those cases, no automated approach could successfully debias the dataset. As we will see in the subsequent experiments, while predictability is a prerequisite for an accurate completion we can largely reduce the bias for a wide set of real-world datasets. This is the case since real-world data is often largely correlated which can be exploited when predicting missing tuples.
 
Moreover, attribute skew as shown in Figure~\ref{fig:syn:red} (lower row) does not seem to have a large influence on the performance of our approach. The reason is that the model can still accurately predict the value of attribute $B$ as long as there is a sufficient amount of training data. 
Finally, as we can see in Figure~\ref{fig:syn:ssar_fan_out} SSAR models are superior over AR models since they can capture fan-out evidence. 
For showing this, we feed the tuples in $T_B$ that share the same tuple in $T_A$ as self-evidence into the SSAR models (which is a type of fan-out evidence as described in Section \ref{sec:schema_models:ssar}).
As we see, if the coherence within the group of tuples in $T_B$ that share a reference to the same tuple in $T_A$ is higher (which we call fan-out predictability) the bias reduction of SSAR compared to AR models improves.

\paragraph{Confidence Intervals.} In addition to bias reduction, we next evaluate the quality of our confidence intervals using synthetic data.
Similar as before, we use a setup with two tables: a complete table $T_A$ with a single attribute $A$ and an incomplete table $T_B$ with a single attribute $B$ where $T_B$ has a foreign-key relationship to $T_A$. Moreover, we vary the predictability as noted in the setup of this experiment. Note that due to a bias, a certain attribute value $b$ of $B$ can appear less/more frequently in the incomplete table compared to the complete table.

\begin{figure}
	\centering
	\includegraphics[width=0.95\linewidth]{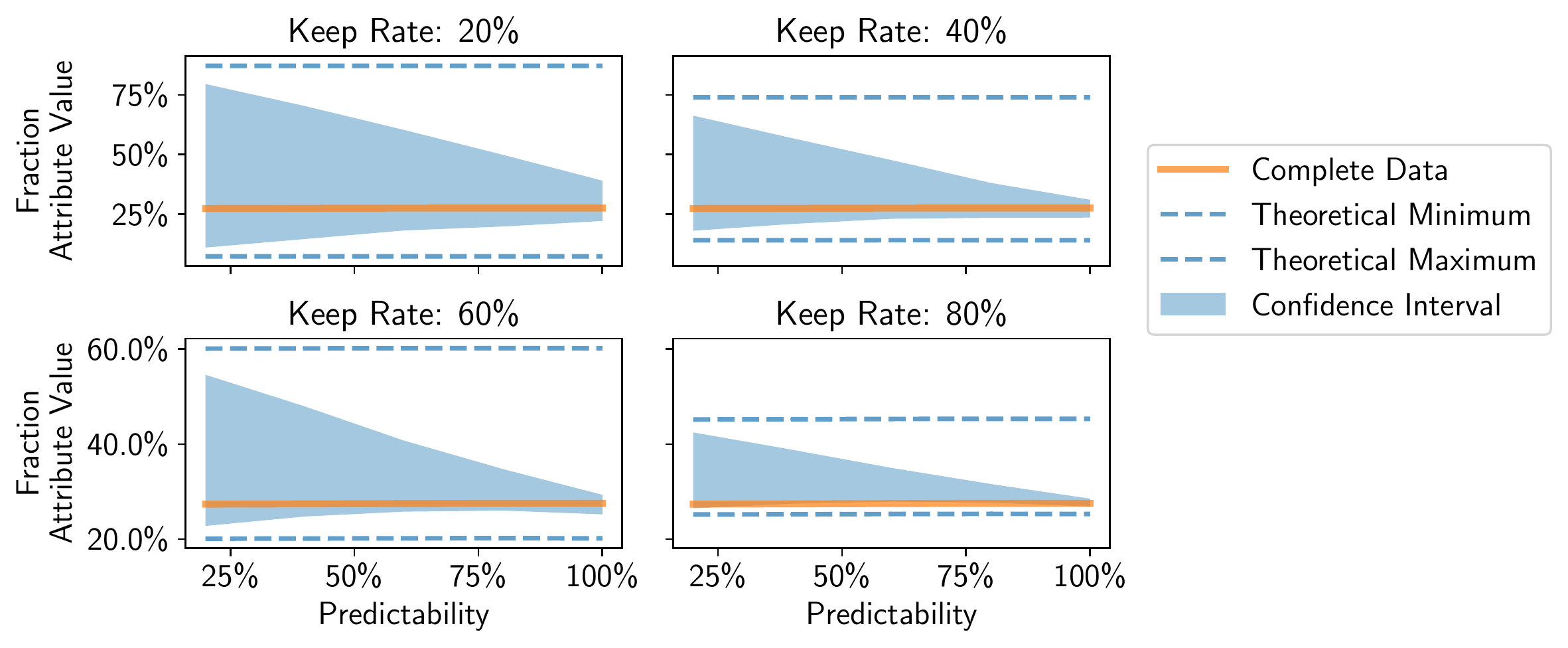}
	\vspace*{-3.5ex}
	\caption{Predicted Confidence Intervals on the Synthetic Data for a Removal Correlation of 40\%. The bounds always capture the true fraction of the attribute value and an increased predictability results in tighter confidence intervals.}
	\vspace*{-5.5ex}
	\label{fig:syn_ci_0.4}
\end{figure}

We now compute the confidence intervals for a count-query over $B$ that reports how often a particular attribute value $b$ occurs. We have chosen the attribute value $b$ with the highest deviation between incomplete and complete data which is a challenging task for \sysname{}. Hence, confidence intervals are particularly of interest.
In Figure \ref{fig:syn_ci_0.4}, we report the fraction of the attribute value $b$ in the true (i.e., original) and the completed database using $95\%$ confidence intervals for the setup described before.

As we can see, the true fraction of the selected attribute value $b$ on the complete dataset is always within the predicted confidence bounds and a larger keep rate results in tighter confidence bounds. 
Moreover, as expected an increased predictability (x-axis) results in more confident completions and thus tighter confidence bounds. 
In addition to the predicted confidence bounds, in Figure \ref{fig:syn_ci_0.4} we also plot the theoretical minimum and maximum of the bounds.
The theoretical minimum and maximum of the bounds can be computed by replacing all respectively none of the missing values with the given attribute value $b$. 
As a sanity check, we see that our confidence bounds also fall into the theoretical bounds. 
In the appendix \ref{sec:app:ext_ci} we present additional results on confidence intervals for the real-world datasets.

\begin{figure*}
	\centering
	\subcaptionbox{Bias Reductions.\label{fig:exp_best_bias}}[0.49\linewidth]{\includegraphics[width=0.99\linewidth]{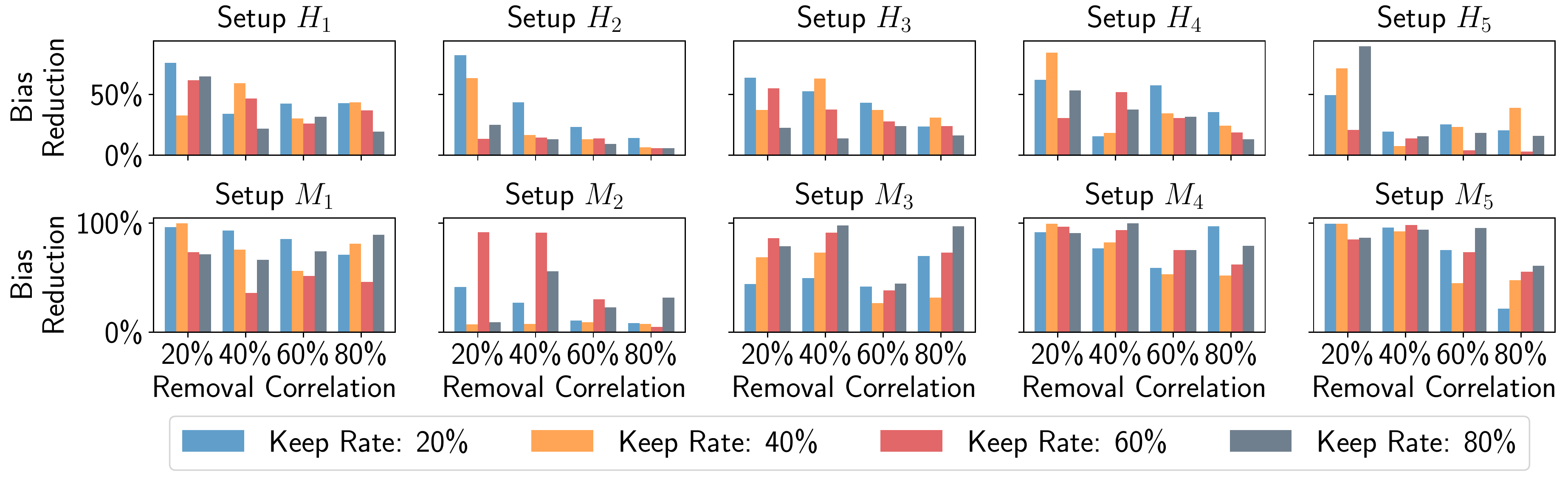}\vspace*{-2ex}}
	\subcaptionbox{Cardinality Corrections.\label{fig:exp_best_count}}[0.49\linewidth]{\includegraphics[width=0.99\linewidth]{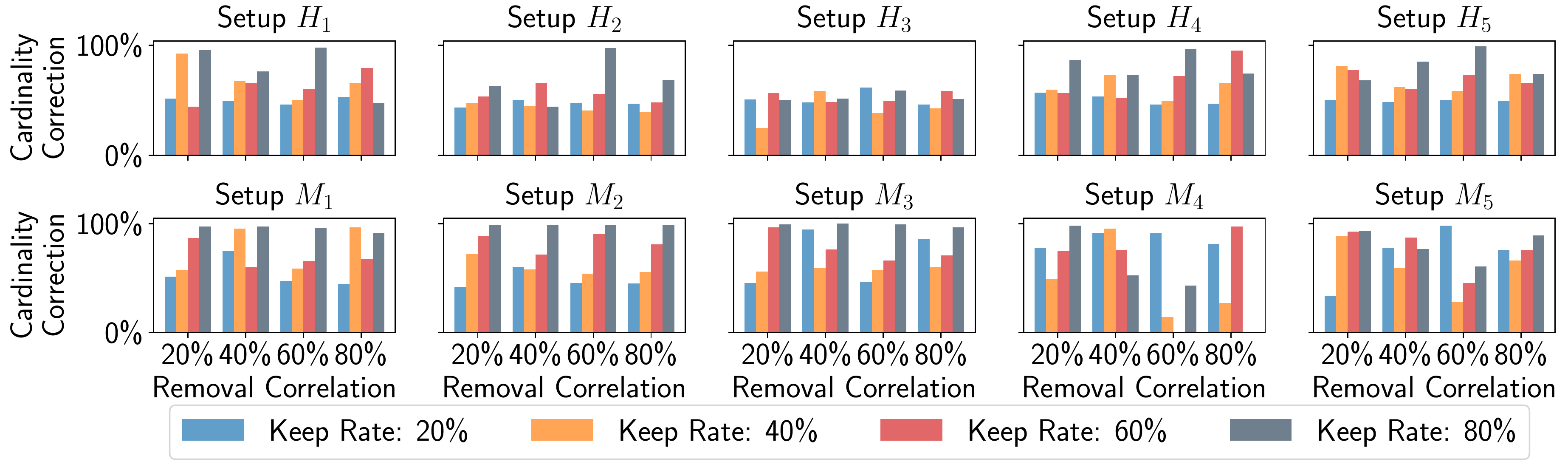}\vspace*{-2ex}}
	\vspace{-3.5ex}
	\caption{Completion Results for Real-World Data using the Setups of the Housing ($H_i$) and Movies ($M_i$) Datasets.}
	\vspace{-2.5ex}
	\label{fig:syn}
\end{figure*}

\subsection{Exp. 2: Data Completion on Real Data}
\label{sec:experiments:exp1}

In this experiment, we analyze how well our approach can complete the two real-world datasets. This is more challenging since the underlying schemas are significantly more complex as depicted in Figures~\ref{fig:datasets:housing} and~\ref{fig:datasets:movies}. Additionally, the data distributions exhibit more interesting correlations.

\paragraph{Completion Setups.} 
Per dataset we have defined five setups as depicted in Table~\ref{fig:datasets:setups} (denoted as $H_i$ and $M_i$ for the housing and movie data, respectively). 
In each setup, we create an incomplete relational dataset by systematically removing tuples using a particular attribute resembling different data types (categorical and continuous) and data distributions.
Similar to the synthetic dataset, we vary the following parameters: \textit{keep rate} and \textit{removal correlation} which are varied from $20\%$ to $80\%$ for all setups. For categorical attributes, we again correlate the removal with the appearance of an attribute value whereas for continuous attributes we correlate it with the normalized attribute value (i.e., to obtain a specific Pearson correlation coefficient).
Moreover, we only keep a small share of all tuple factors - $20\%$ for the movie dataset and $30\%$ for the housing dataset to compensate for an overall smaller dataset.
In addition, to include some even more challenging setups for the movies dataset we additionally remove all tuples in the $m:n$ relationship tables (i.e., \texttt{movie\_company} etc.) which do not have a matching tuple after the removal. For the setups $M_4$ and $M_5$ we additionally remove $20\%$ of the \texttt{movie} tuples.

\paragraph{Results.} As discussed before, an interesting metric is how well we could debias the incomplete data using our completion models under the different setups. The results are shown Figure~\ref{fig:exp_best_bias} for all five setups given a variety of keep rates (between $20\%$ and $80\%$) and removal correlations. As we see, the bias can significantly be reduced for all setups indicating the high quality of our completion models. This especially holds for the setups of the movies dataset where up to $100\%$ of the bias can be removed. In general, a lower removal correlation is beneficial for our approach. The reason is that the lower the correlation, the more examples of high attribute values (for continuous attributes) remain in the training set and thus the model can learn more precisely what leads to those higher values. During the completion it can then predict more accurately whether larger values are likely to occur. The keep rates do not seem to have a significant impact. The reason is that there are two opposing effects. On the one hand, a larger keep rate leads to a larger training dataset and the model can thus learn the distribution more accurately. On the other hand, the absolute error $|\mathit{AVG}_{\mathit{complete}}(X)-\mathit{AVG}_{\mathit{incomplete}}(X)|$ becomes smaller and the model has to predict more extreme values to correct the bias. Consequently, we do not see more accurate completions for larger keep rates. 

However, the quality of the completion varies for the different setups. The reason is that the remaining evidence, i.e., the complete tables in the schema are not equally useful. Some attributes of available data are in general less predictable and if those are used for a biased removal, it becomes harder to correct the bias. Interestingly, we do not see the general trend that the completions become less accurate for longer completion paths. Recall that for setups $M_4$ and $M_5$, all single completion paths span at least five tables. However, the completions are significantly more accurate than those of $M_2$. This highlights that the predictability of the biased attribute has the most significant impact on the bias reduction. 
In general, for setups such as $H_2$ and $M_2$ where the evidence of the complete tables does not allow an accurate prediction of the biased attribute, the models cannot correct the bias. This is consistent with our findings on synthetic data.

\paragraph{Count Correction.} We are also interested in how accurately the table sizes are estimated using different ratios of available tuple factors. Similarly to the bias reduction we define the \textit{cardinality correction} as $1-\frac{|\mathit{Completed\,Tuples}|-|\mathit{Complete\,Tuples}|}{|\mathit{Incomplete\,Tuples}|-|\mathit{Complete\,Tuples}|}.$
As we can see in Figure~\ref{fig:syn}, the cardinalities of the complete tables can relatively accurately be predicted even though only $20-30\%$ of all tuple factors are kept in the incomplete datasets.

\begin{table*}
	\scriptsize
	\centering
	\caption{Queries used for Figure~\ref{fig:exp_qp} with both Joins and complex Filter Predicates, Aggregations and Groupings.}\label{tab:queries}
	\vspace*{-2.5ex}
	\begin{tabular}{llll}\toprule
		Dataset & Setup & Query & SQL  \\\midrule
Housing & $H_1$ & $Q_{1}$ & \texttt{SELECT SUM(price) FROM apartment WHERE room\_type='Entire home/apt';} \\
		Housing & $H_2$ & $Q_{2}$ & \texttt{SELECT COUNT(*) FROM apartment WHERE room\_type='Entire home/apt' AND property\_type='House' GROUP BY property\_type;} \\
		Housing & $H_3$ & $Q_{3}$ & \texttt{SELECT COUNT(*) FROM apartment WHERE property\_type='House';} \\
		Housing & $H_4$ & $Q_{4}$ & \texttt{SELECT COUNT(*) FROM landlord WHERE landlord\_since$\ge$2011;} \\
		Housing & $H_5$ & $Q_{5}$ & \texttt{SELECT AVG(landlord\_response\_rate) FROM landlord WHERE landlord\_response\_time$\ge$2;} \\
		Housing & $H_1$ & $Q_{6}$ & \texttt{SELECT AVG(price) FROM landlord NATURAL JOIN apartment WHERE room\_type='Entire home/apt' GROUP BY landlord\_since;} \\
		Housing & $H_2$ & $Q_{7}$ & \texttt{SELECT COUNT(*) FROM landlord NATURAL JOIN apartment WHERE accommodates$\ge$3 GROUP BY landlord\_since;} \\
		Housing & $H_3$ & $Q_{8}$ & \texttt{SELECT COUNT(*) FROM landlord NATURAL JOIN apartment WHERE landlord\_since$\ge$2013 GROUP BY landlord\_since;} \\
		Housing & $H_4$ & $Q_{9}$ & \texttt{SELECT SUM(landlord\_since) FROM landlord NATURAL JOIN apartment WHERE room\_type='Entire home/apt' AND landlord\_response\_time$\ge$2;} \\
		Housing & $H_5$ & $Q_{10}$ & \texttt{SELECT AVG(landlord\_response\_rate) FROM landlord NATURAL JOIN apartment WHERE room\_type='Entire home/apt' AND landlord\_response\_time$\ge$2;} \\
		Movies & $M_1$ & $Q_{1}$ & \texttt{SELECT COUNT(*) GROUP BY production\_year;} \\
		Movies & $M_2$ & $Q_{2}$ & \texttt{SELECT COUNT(*) FROM movie WHERE genre='Drama' GROUP BY production\_year;} \\
		Movies & $M_3$ & $Q_{3}$ & \texttt{SELECT COUNT(*) FROM movie WHERE genre='Drama' GROUP BY country;} \\
		Movies & $M_4$ & $Q_{4}$ & \texttt{SELECT AVG(birth\_year) FROM director WHERE gender='m';} \\
		Movies & $M_5$ & $Q_{5}$ & \texttt{SELECT COUNT(*) FROM company WHERE country\_code='[us]';} \\
		Movies & $M_1$ & $Q_{6}$ & \texttt{SELECT SUM(production\_year) FROM movie NATURAL JOIN movie\_director NATURAL JOIN director WHERE birth\_country='USA' GROUP BY production\_year;} \\
		Movies & $M_2$ & $Q_{7}$ & \texttt{SELECT COUNT(*) GROUP BY country\_code;} \\
		Movies & $M_3$ & $Q_{8}$ & \texttt{SELECT COUNT(*) FROM movie NATURAL JOIN company NATURAL JOIN movie\_companies WHERE country\_code='[us]' GROUP BY production\_year;} \\
		Movies & $M_4$ & $Q_{9}$ & \texttt{SELECT COUNT(*) FROM movie NATURAL JOIN movie\_director NATURAL JOIN director WHERE gender='m';} \\
		Movies & $M_5$ & $Q_{10}$ & \texttt{SELECT COUNT(*) FROM movie NATURAL JOIN company NATURAL JOIN movie\_companies WHERE country\_code='[us]' GROUP BY country;} \\
		\bottomrule
	\end{tabular}
\end{table*}

\begin{figure*}
	\centering
	\vspace*{-0.5ex}
	\includegraphics[width=\linewidth]{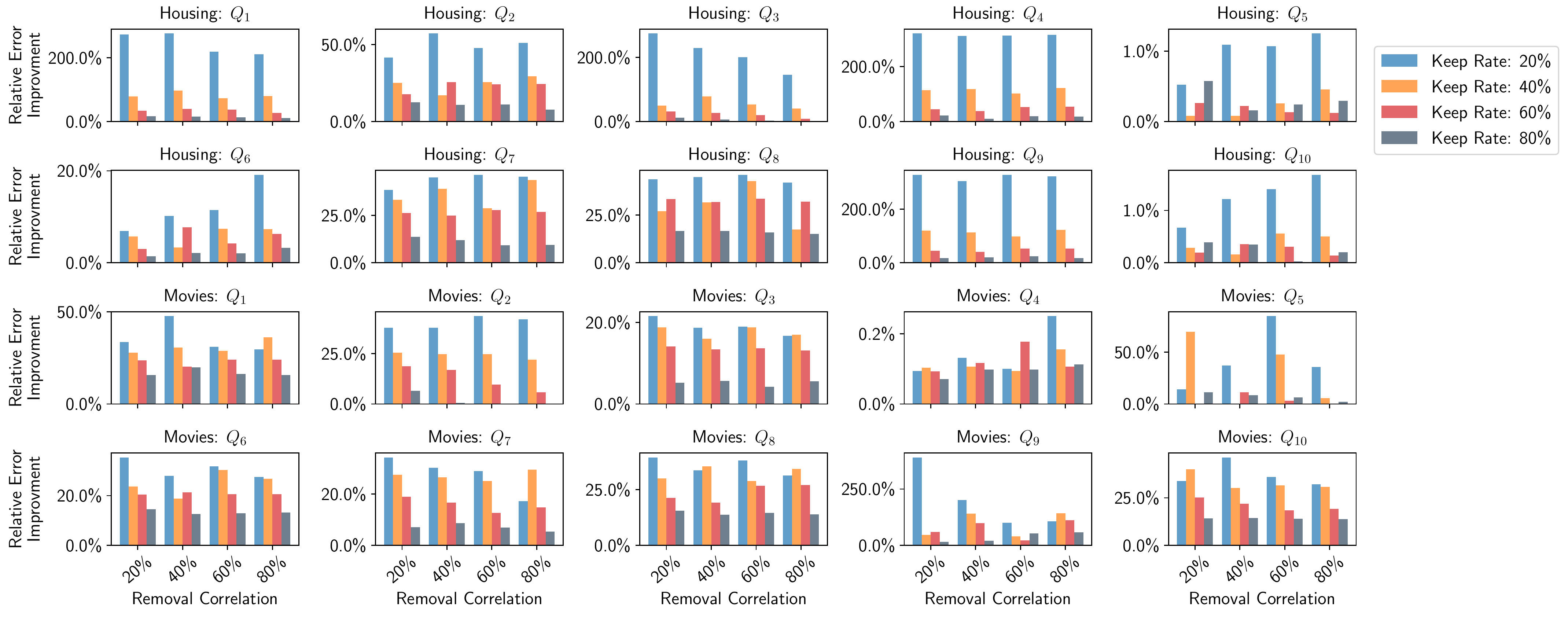}
	\vspace*{-3.5ex}
	\caption{Improvement of Average Relative Error due the Completion (i.e., higher is better). Improvements for individual \texttt{COUNT}, \texttt{AVG} and \texttt{SUM} queries are shown as separate plots.}
	\vspace*{-2.5ex}
	\label{fig:exp_qp}
\end{figure*}

\subsection{Exp. 3: Query Processing}
\label{sec:experiments:query_processing}

\paragraph{Completion Setups.} We now investigate the end-to-end performance of our approach for query processing. To this end, we use a workload of both single table and join queries with aggregates and various filter predicates (cf. Table~\ref{tab:queries}). We then derive incomplete datasets similar to Exp. 1. and compare the relative error of the queries computed on the incomplete dataset and our completed dataset (using the original complete datasets as ground truth). We show the absolute improvement for the relative error for the queries. 

\begin{figure*}
\centering
\includegraphics[width=0.85\linewidth]{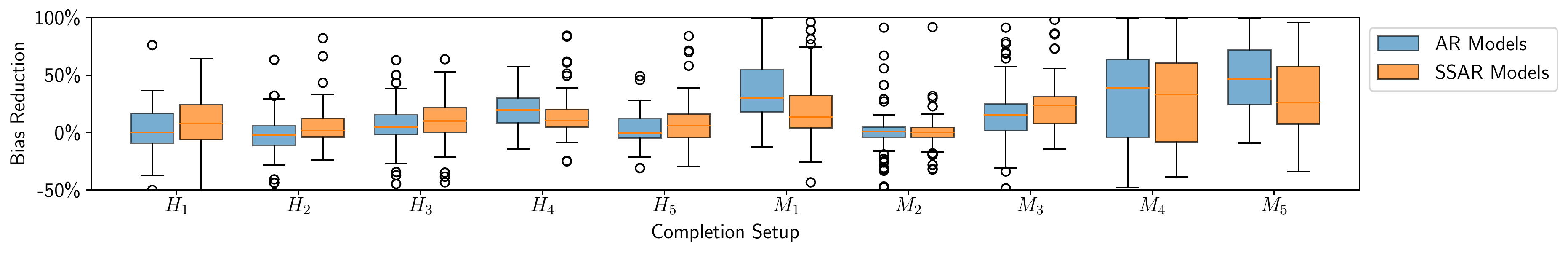}
\vspace*{-3.5ex}
\caption{AR vs. SSAR Models. No clear advantage of one model can be derived demonstrating the need for model selection.}
\vspace*{-2.5ex}
\label{fig:exp_ar_vs_ssar}
\end{figure*}

\begin{figure*}
	\centering
	\includegraphics[width=0.85\linewidth]{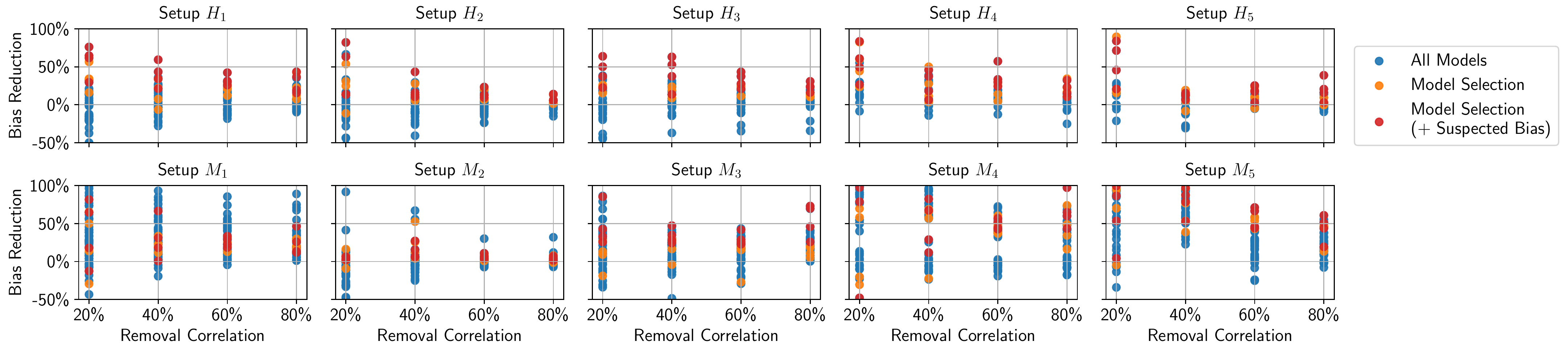}
	\vspace*{-3.5ex}
	\caption{Quality of Models selected by our Model Selection Strategy vs. all Models for the different Housing Setups ($H_1$ to $H_5$) and Movies Setups ($M_1$ to $M_5$). We can select the best performing model in almost all cases if a bias is suspected. If this information is not available, the selected model is usually among the most suitable models for the completion.}
	\label{fig:exp_model_selection}
	\vspace*{-3.5ex}
\end{figure*}

\begin{figure}
	\centering
	\includegraphics[width=0.8\columnwidth]{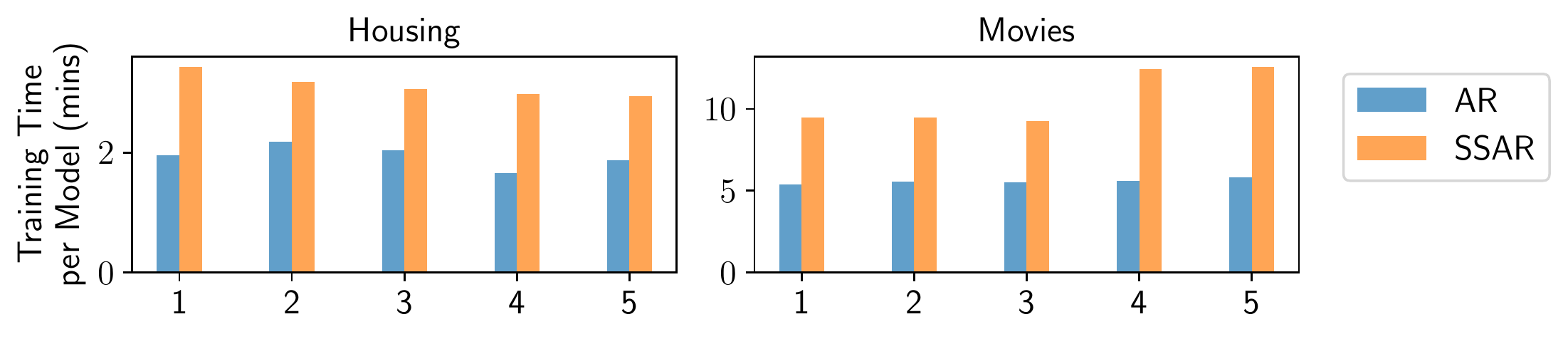}
	\vspace*{-3.5ex}
	\caption{Time required for Training.}
	\label{fig:exp_training_time}
	\vspace*{-2.5ex}
\end{figure}

\begin{figure}
	\centering
	\includegraphics[width=0.95\columnwidth]{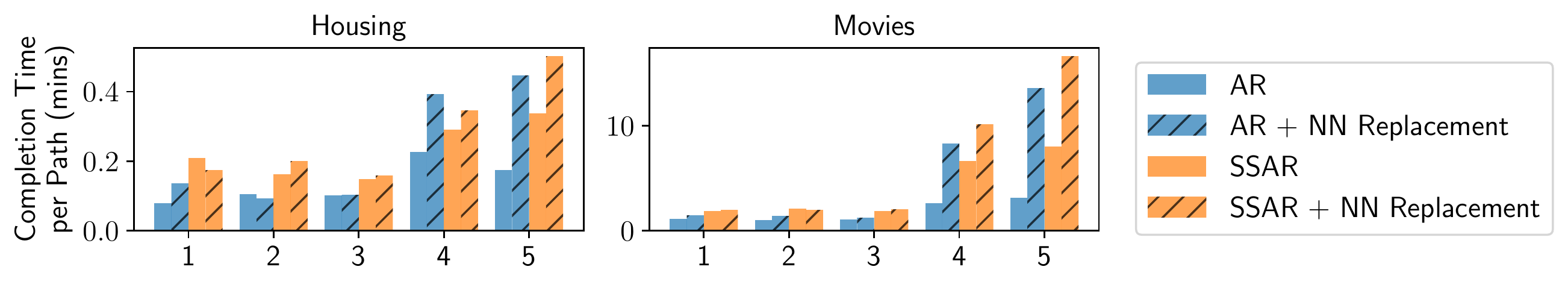}
	\vspace*{-3.5ex}
	\caption{Time required for completing one Path.}
	\label{fig:exp_query_time}
	\vspace*{-3.5ex}
\end{figure}

\paragraph{Results.} As we can see in Figure~\ref{fig:exp_qp}, we can achieve significant improvements motivating the use of our approach for practical applications. We can see that \texttt{COUNT} and \texttt{SUM} queries are in general largely improved while the improvements for the \texttt{AVG} queries are smaller. The reason is that for \texttt{AVG} queries the improvement depends on the scaling and translation of the attribute as well as the absolute error introduced by the biased removal. This varies largely for the different attributes in our datasets. This emphasizes the importance of the bias reduction metric in the first experiment. 

In addition, we noticed that for join queries on the smaller housing dataset and low keep rates the predictions of our models tend to be inferior to the incomplete dataset. In this case, the AR and SSAR models cannot observe sufficient training data to make accurate predictions. We thus recommend not to use our approach if the number of available tuples is very low.
However, for larger datasets it is also more time-consuming to complete them manually and in these cases our approach achieves significantly more accurate query results as we can see from Figure~\ref{fig:exp_qp}.

\vspace{-0.5ex}\subsection{Exp. 4: Accuracy and Performance Aspects}
\label{sec:experiments:performance}

\paragraph{Model and Path Selection.} We next investigate how reliable our model and path selection works. To this end, we plot all bias reductions and the performance of the model selection strategies in Figure~\ref{fig:exp_model_selection}. 
If we provide the information which bias is suspected in the data (red dots in Figure~\ref{fig:exp_model_selection}), we often pick the optimal path and model. However, even if this information is not available (orange dots in Figure~\ref{fig:exp_model_selection}), we select models that can effectively reduce bias. 

\paragraph{Training Time.} In Figure~\ref{fig:exp_training_time} we depict the average training time of the AR and SSAR models for the different completion setups. As we can see in general AR models require less training time ($<2$ minutes for housing and $<6$ minutes for the movies dataset). The reason is twofold. First, the models do not require acyclic walks on the schema which have to be performed to gather training data for the SSAR models. Moreover, the models are not as complex since the tree models for the schema walks are not required. While SSAR models require a longer training time, this can be justified with a better performance for some completion setups. 

\paragraph{Completion Time.} Finally, we discuss the time needed for data completion. As we can see in Figure~\ref{fig:exp_query_time} the completion via one path takes less than $30$ seconds for all setups of the housing dataset. For the larger movies dataset, however, the completion took less than two minutes for the completion setups $M_1-M_3$. For the more challenging setups with long-distance completion paths (distance of four) the completion takes around $16$ minutes. However, here millions of tuples have to be synthesized. 
Moreover, we can see that the nearest neighbor replacement increases the runtime of the completion. 
As mentioned before, for those scenarios we can alternatively generate the data offline. 

\paragraph{SSAR vs. AR Model.} First, we want to compare the performances of AR and SSAR completion models. Recall that the SSAR models obtain additional fan-out evidence. We have depicted the distributions of all bias reductions in Figure~\ref{fig:exp_ar_vs_ssar}. As we can see neither AR nor SSAR models always outperform the other class of models. Instead, it again depends on the concrete setup we are considering. This motivates the model and path selection algorithm which aims at identifying such cases and chooses alternative models.

 \vspace{-2.5ex}\section{Related Work}
\label{sec:related_work}

\paragraph{Missing Data in OLAP.} Closest to our approach is probably the recent Themis \cite{orr2020themis} system. Different from \sysname{}, Themis is restricted to work for a single table and requires aggregate information. Themis either reweights existing tuples or learns probabilistic models for missing groups. The techniques for leveraging aggregate knowledge such as iterative proportional fitting could seamlessly be integrated in our approach.
Chung et al. \cite{chung2016unknowns} estimate the impact of missing tuples on aggregate queries when several data sources are integrated by observing reoccurring tuples. While \sysname{} similarly helps for the case of different data sources with varying quality again only the single table case is discussed here.
There has also been work on determining when incomplete data still leads to complete query results \cite{motro1989integrity, levy1996obtaining} or which parts of the result are complete \cite{lang2014partial} which is orthogonal to our work.

\paragraph{Data Generation.} 
In order to compensate missing tuples, we synthesize missing data using AR and SSAR models. This is related to approaches that synthesize tuples \cite{sun2015bayesian, fan2020relationalsynthesis, ijcai2019-287, DBLP:journals/corr/abs-1811-11264} using deep models such as GANs \cite{NIPS2014_5423, NIPS2016_6125}. A main motivation is to synthesize data satisfying data privacy.
In contrast to \sysname, the models typically only support individual tables instead of complex schemas.

\paragraph{Uncertain and Probabilistic Databases.}
Another line of work \cite{feng2019uncertainty, DBLP:conf/icdt/SundarmurthyKLN17} uses the possible world semantics \cite{10.1145/38713.38724} to handle uncertain data, i.e., either tuple values or the inclusion of tuples in the dataset are uncertain. The goal is to estimate possible results for queries.
Alternatively, uncertainty can be modeled using probabilistic databases \cite{Wang:2008:BML:1453856.1453896, jha2012probabilistic, Rekatsinas:2012:LSD:2213836.2213879,Sen:2009:PME:1644245.1644258, Dalvi:2007:EQE:1285882.1285906, Olteanu2009Sprout, suciu2010efficient} where tuples or sets of tuples are annotated with probabilities. 
In contrast to our work, missing tuples cannot be handled directly. Possibly missing tuples would have to be manually inserted in the database and annotated with a probability which is challenging since the user often does not have an understanding of what data is missing.

\paragraph{Data Cleaning.} Our approach is also related to data cleaning.
A major direction in data cleaning are approaches for value imputation \cite{10.1145/2882903.2912574}. For value imputation, there exist many techniques that leverage probabilistic graphical models \cite{DBLP:journals/pvldb/RekatsinasCIR17}, relational dependency networks \cite{mayfield2010eracer} or neural approaches \cite{mlsys2020_123, pmlr-v80-yoon18a}. All these approaches, however, cannot synthesize completely missing tuples as we do. 
Another interesting direction, is \cite{wang2014sampleclean} which estimates the result of aggregate queries by cleaning a sample of dirty data. However, again missing tuples are not being compensated for.

 \balance{}
\vspace{-1.5ex}\section{Conclusion and Future Work}
\label{sec:conclusion}

In this paper, we have introduced \sysname{} --- an approach that approximates queries over a relational database in cases where only incomplete data is available (i.e., tuples in individual tables are missing). 
In our experimental evaluation, we have demonstrated that our approach can synthesize missing data with high accuracy and thus enables improved decision making on top of incomplete relational databases. 
In future work, we also want to investigate how the models devised can be used for tasks like missing data imputation or other downstream tasks (e.g., learning a classification model) that can now use the completed dataset as input. In addition, we believe that combining our approach with probabilistic databases is also a promising direction.

\vspace{-1.5ex}\section{Acknowledgments}

This research was partly funded by the BMBF Project \emph{KompAKI}, the Hochtief project \emph{AICO} (AI in Construction) as well as the HMWK cluster project \emph{3AI} (The Third Wave of AI).  
\bibliographystyle{abbrv}
\bibliography{bib}

\newpage{}
\appendix
\section{Extended Confidence Interval Experiments}
\label{sec:app:ext_ci}

\begin{figure*}
	\centering
	\includegraphics[width=0.8\linewidth]{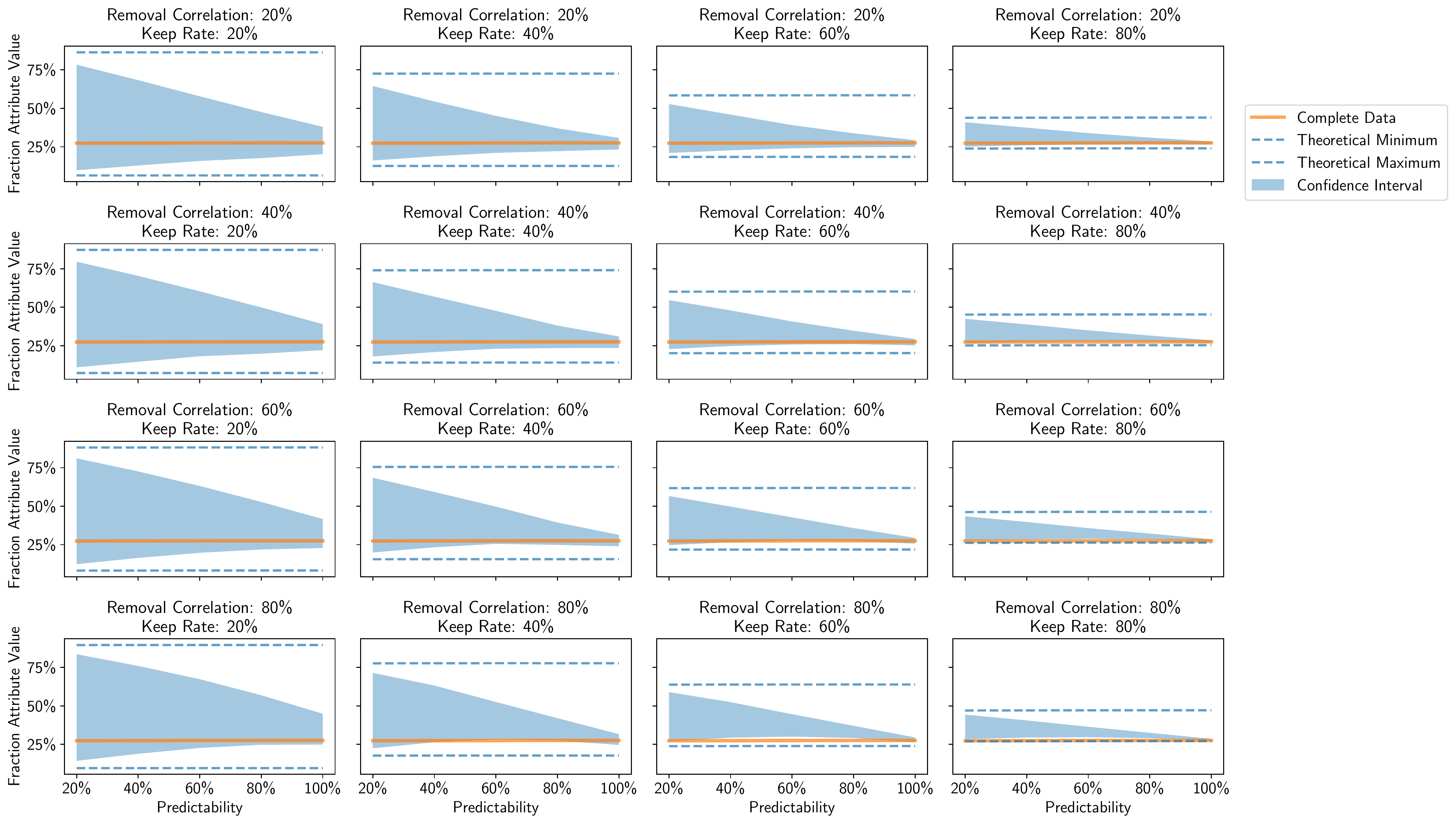}
	\vspace*{-3.5ex}
	\caption{Confidence Intervals for all Synthetic Data Setups. An increased predictability results in tighter estimated confidence intervals.}
	\vspace*{-2.5ex}
	\label{fig:syn_ci}
\end{figure*}

\begin{figure*}
	\centering
	\includegraphics[width=0.8\linewidth]{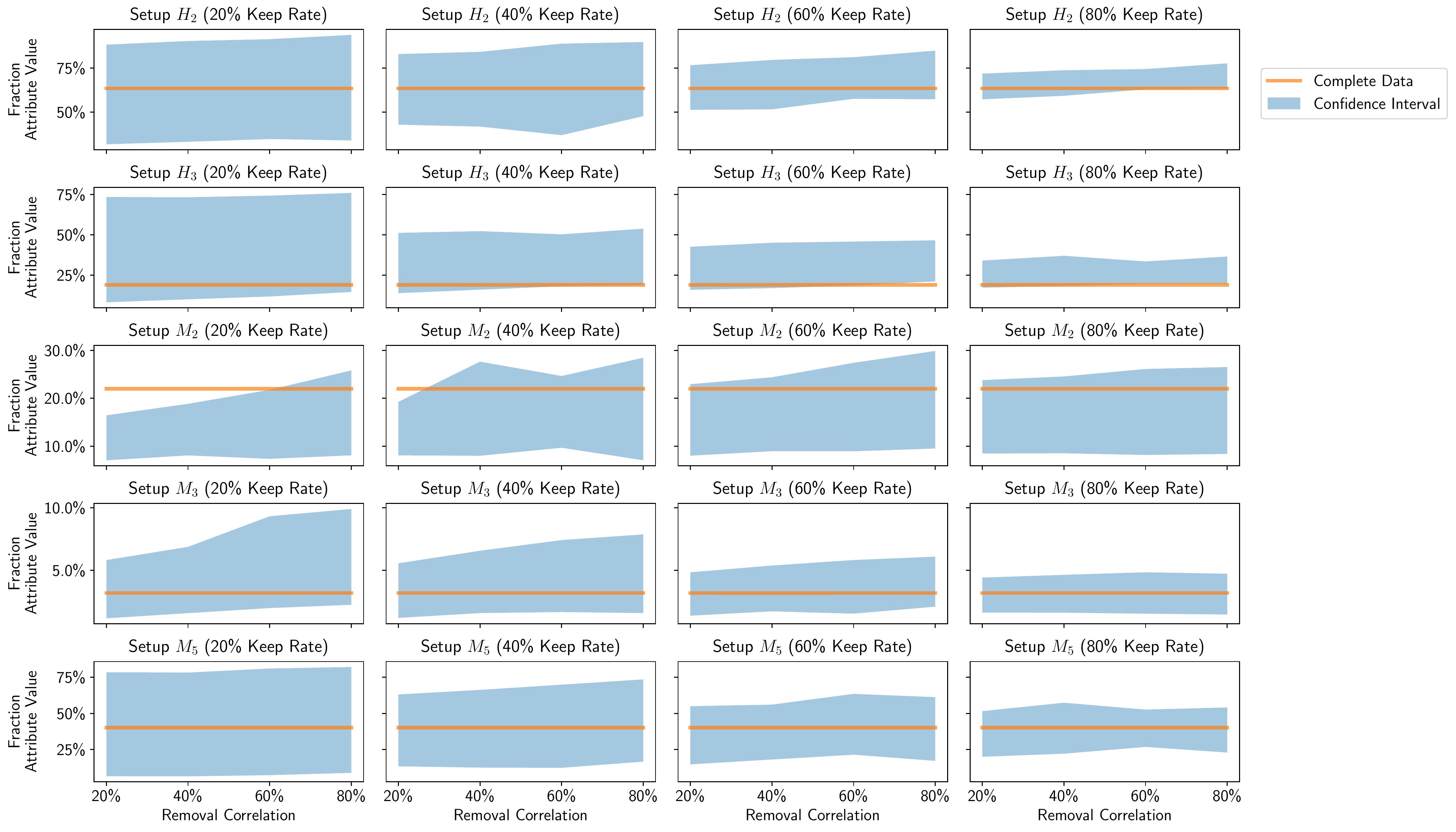}
	\vspace*{-2.5ex}
	\caption{Confidence Intervals for all Categorical Attributes for the real-world Datasets. Note that we show the confidence intervals w.r.t. the removal correlation on the x-axis since for real-world data the predictability cannot be controlled explicitly.}
	\vspace*{-2.5ex}
	\label{fig:real_ci}
\end{figure*}

In the following, we present further results of using our confidence estimates that we could not add to the revision due to space restrictions as mentioned before.
The results of the additional experiments for synthetic and real world data sets (i.e., housing and movies) are depicted in Figures~\ref{fig:syn_ci} and \ref{fig:real_ci}, respectively.

\underline{Synthetic data.} We first use the synthetic dataset, since for this dataset we can explicitly control how accurately the true share of missing tuples can be predicted given the evidence tuples using the \emph{predictability} parameter (i.e., how much the missing data correlates with the available data).
For this experiment, we use the same setup with two tables as for Exp. 1 in the paper: a complete table $T_A$ with a single attribute $A$ and an incomplete table $T_B$ with a single attribute $B$ where $T_B$ has a foreign-key relationship to $T_A$ and we vary the predictability as noted in the setup of this experiment. Note that due to a bias, a certain attribute value $b$ of $B$ can appear less /more frequently in the incomplete table as in the true (complete) database.

We now compute the confidence intervals for a count-query over $B$ that reports how often a particular attribute value $b$ occurs. We have chosen the attribute value $b$ with the highest deviation between incomplete and complete data which is in particular challenging for \sysname{}. Hence, confidence intervals are particularly of interest.
The $95\%$ confidence intervals for the count-query mentioned before for varying parameters can be seen in Figures~\ref{fig:syn_ci}.
In all plots, we can compare the fraction of the selected attribute value $b$ in the complete (true) database (orange line) and the confidence intervals (blue area) that our approach computes for the completion. 
As we can see in Figure~\ref{fig:syn_ci}, when increasing the predictability for the missing data (i.e., attribute $B$), the resulting confidence intervals are becoming indeed tighter since the models can more confidently synthesize the $B$ values of the missing tuples. Moreover, a larger \emph{keep rate} (i.e., fewer missing tuples in $T_B$) also results in tighter confidence intervals as expected.

In addition to the predicted confidence intervals, we also plot the theoretical minimum and maximum of the bounds.
The theoretical minimum and maximum can be computed by replacing all respectively none of the missing values with the given attribute value $b$ of $B$ we used for the plot. 
As a sanity check, we see that our confidence bounds also fall into the theoretical bounds.

Another important observation when looking at  Figure~\ref{fig:syn_ci} is that the true fraction is in (almost) all cases within our confidence intervals.
However, in some rare cases the true fraction is slightly lower than the predicted lower bound of the confidence intervals 
(e.g., for a keep rate of 80\%).
In these cases, the correlations of missing tuples deviate too much from the correlations of the training data. This can happen if we only remove a few (extreme) outlier tuples from the complete data to create the incomplete database. Note that a similar effect that the true value is out of the confidence bounds can also be observed for confidence intervals in other applications (e.g., approximate query processing) when some (extreme) outliers are missing in the sample but occur in the true (complete) database.

\underline{Real-world data.}  In addition to synthetic data, we also evaluated the confidence intervals on real-world data.
Here we used the setups as listed in Table~\ref{fig:datasets:setups} in the paper. In each of those setups, we investigate how well the values of different attributes of missing tuples in different real-world datasets (i.e., housing and movies) can be restored (e.g., the room type for setup $H_2$). 

In this experiment, we first concentrated on the categorical attributes such as the room type mentioned before. Similar to the synthetic data, for categorical attributes a certain attribute value appears less/more frequently in the incomplete data due to the bias. We thus issue a count-query over the attribute that reports, how often a particular attribute value occurs. We again select the attribute value, at which the deviation between incomplete and complete data and thus the error of the query is maximized, e.g., a certain apartment type that occurs much more frequently in the complete dataset.
The results for different setups (i.e., different attributes) are given in Figure~\ref{fig:real_ci}. As we can see, as before for the synthetic data in nearly all cases on real data the true share of the restored attribute value (in the true complete database) is either contained in the confidence intervals or close to the predicted bounds. 

Note that in contrast to the synthetic datasets, we cannot vary the predictability since correlations are given by the data. 
Instead, as in other experiments with real-world data (e.g., Exp. 2) we varied the removal correlation (as shown on the x-axis of Figure~\ref{fig:real_ci}). Note that as expected this parameter does not have a clear influence on the tightness of the completion intervals. The reason is that this parameter controls the intensity of the bias in the incomplete table but for a constant predictability, i.e., for a larger removal correlation fewer tuples with a particular attribute value remain in the incomplete dataset. 
Instead, it is important that the confidence intervals (blue area) contain the true fractions (orange line) in most cases again. Moreover, we also repeated the experiment for continuous attributes (e.g., rental prices of apartments) and observed comparable results.

\end{document}